\documentclass[12pt]{article}

\usepackage{scicite}
\usepackage{soul}
\usepackage{times}
\usepackage{url}
\usepackage{comment}
\usepackage[normalem]{ulem}
\useunder{\uline}{\ul}{}
\usepackage{lscape}
\usepackage{amsmath}
\usepackage{amssymb}
\usepackage{booktabs}
\usepackage{graphicx}
\usepackage{upgreek}
\usepackage{textcomp}
\usepackage{float}
\usepackage{pdfpages}
\usepackage[noadjust]{cite}

\usepackage{xcolor}

\definecolor{goldenpoppy}{rgb}{0.99, 0.76, 0.0}
\definecolor{mygreen}{rgb}{0.19,0.55,0.11}

\newenvironment{sciabstract}{%
\begin{quote} \bf}
{\end{quote}}

\newcounter{lastnote}

\newcommand{\E}{PSR~J0514$-$4002E}
\newcommand{\Msun}{\mathrm{M}_\odot}
\newcommand{\Rsun}{\mathrm{R}_\odot}

\title{A pulsar in a binary with a compact object in the mass gap between neutron stars and black holes}

\author
{Ewan~D.~Barr,$^{1\ast\dagger}$
Arunima Dutta,$^{1\ast\dagger}$
Paulo~C.~C.~Freire,$^{1}$
Mario Cadelano,$^{3,4}$\\
Tasha Gautam,$^{1}$
Michael Kramer,$^{1}$
Cristina Pallanca,$^{3,4}$
Scott~M.~Ransom,$^{9}$\\
Alessandro Ridolfi,$^{1,7}$
Benjamin~W.~Stappers,$^{2}$
Thomas~M.~Tauris,$^{1,13}$\\
Vivek Venkatraman Krishnan,$^{1}$
Norbert Wex,$^{1}$
Matthew~Bailes,$^{5,6}$\\
Jan~Behrend,$^{1}$
Sarah Buchner,$^{8}$
Marta~Burgay,$^{7}$
Weiwei~Chen,$^{1}$\\
David~J.~Champion,$^{1}$
C. -H.~Rosie~Chen,$^{1}$
Alessandro Corongiu,$^{7}$\\
Marisa Geyer,$^{8,12\ddagger}$
Y.~P.~Men,$^{1}$
Prajwal~V.~Padmanabh,$^{1,10,11}$
Andrea Possenti$^{7}$\\
\\
\normalsize{$^{1}$ Max-Planck-Institut f\"{u}r Radioastronomie, 53121 Bonn, Germany}\\
\normalsize{$^{2}$ Jodrell Bank Centre for Astrophysics, Department of Physics and Astronomy,}\\
\normalsize{The University of Manchester, Manchester, M13 9PL, UK}\\
\normalsize{$^{3}$ Dipartimento di Fisica e Astronomia “Augusto Righi'', Universit\`a degli Studi di Bologna,}\\
\normalsize{40129 Bologna, Italy}\\
\normalsize{$^{4}$ Istituto nazionale di astrofisica-Osservatorio di Astrofisica e Scienze dello Spazio di Bologna,}\\
\normalsize{I-40129 Bologna, Italy}\\
\normalsize{$^{5}$ Centre for Astrophysics and Supercomputing, Swinburne University of Technology,}\\
\normalsize{PO Box 218 Hawthorn, Vic, 3122, Australia}\\
\normalsize{$^{6}$ Australian Research Council Centre of Excellence for Gravitational Wave Discovery (OzGrav)}\\
\normalsize{$^{7}$ Istituto nazionale di astrofisica}-Osservatorio Astronomico di Cagliari, I-09047, Selargius, Italy\\
\normalsize{$^{8}$ South African Radio Astronomy Observatory, Observatory 7925, South Africa}\\
\normalsize{$^{9}$ National Radio Astronomy Observatory, Charlottesville, VA, 22903, USA}\\
\normalsize{$^{10}$ Max Planck Institute for Gravitational Physics (Albert Einstein Institute),}\\
\normalsize{D-30167 Hannover, Germany}\\
\normalsize{$^{11}$ Leibniz Universit\"{a}t Hannover, D-30167 Hannover, Germany}\\
\normalsize{$^{12}$ Department of Astronomy, University of Cape Town, Rondebosch, Cape Town, 7700,}\\
\normalsize{South Africa}\\
\normalsize{$^{13}$ Department of Materials and Production, Aalborg University}\\
\normalsize{DK-9220~Aalborg {\O}st, Denmark}\\
\normalsize{$^\ast$Corresponding author;}\\
\normalsize{E-mail: ebarr@mpifr-bonn.mpg.de, adutta@mpifr-bonn.mpg.de}\\
\small{$\dagger$These authors contributed equally to this work.}\\
\small{$\ddagger$Present address: High Energy Physics, Cosmology \& Astrophysics Theory (HEPCAT) Group,}\\
\small{Department of Mathematics \& Applied Mathematics, University of Cape Town,}\\
\small{Cape Town 7700, South Africa}
}

\date{}

\begin{document} 

\baselineskip 24pt

\maketitle 

\begin{sciabstract}

Among the compact objects observed in gravitational wave merger events a few have masses in the gap between the most massive neutron stars (NSs) and least massive black holes (BHs) known. Their nature and the formation of their merging binaries are not well understood. We report on pulsar timing observations using the Karoo Array Telescope (MeerKAT) of \E, an eccentric binary millisecond pulsar in the globular cluster NGC~1851 with a total binary mass of $3.887 \pm 0.004$ solar masses ($\Msun$). The companion to the pulsar is a compact object and its mass (between $2.09$ and $2.71 \, \Msun$, 95\% confidence interval) is in the mass gap, so it either is a very massive NS or a low-mass BH. We propose the companion was formed by a merger between two earlier NSs.

\end{sciabstract}

Globular clusters (GCs) are dense, gravitationally bound, stellar clusters. They have been observed to host a large number of low-mass X-ray binaries (LMXBs) consisting of a compact object accreting material from a donor star. LMXBs are $\sim 10^3$ times more abundant per unit of stellar mass in GCs than in the disk of the Milky Way galaxy (Galactic plane) \cite{1975ApJ...199L.143C}. This is due to the high stellar densities at the centre of GCs, which increase the rate of exchange encounters in which neutron stars (NSs) acquire low-mass main sequence (MS) companions. The MS stars evolve until they start transferring mass to the NS, at which point an LMXB is formed.

These X-ray binaries are expected to produce millisecond pulsars (MSPs, radio-emitting neutron stars with spin periods $P < 10$ ms) in almost circular orbits around low-mass companions\cite{1982Natur.300..728A,2023pbse.book.....T}. There are a total of 305 pulsars known in 40 GCs\cite{gc_pulsar_list_url}, the vast majority of which are MSPs. Most of the systems in GCs are similar to the MSP population found in the Galactic plane, although their orbital eccentricities are often higher, which is thought to be a result of close encounters with other stars\cite{1992RSPTA.341...39P}.

In GCs with the densest cores, any particular star - or MSP - is likely to experience multiple exchange encounters over its lifetime\cite{2014A&A...561A..11V}. A possible outcome is the exchange of a low-mass companion of an MSP for either a massive white dwarf (WD) or another NS, resulting in a massive, eccentric MSP binary  \cite{1991ApJ...374L..41P,2004ApJ...606L..53F,2012ApJ...745..109L,2021MNRAS.504.1407R}. Observing such systems allows their component masses to be measured and can test theories of gravity\cite{2006ApJ...644L.113J}. The same process could also produce a  MSP--black hole (BH) system \cite{2018PhRvL.120s1103K}.

\section*{The millisecond pulsar binary \E}

A survey searching for MSPs in GCs \cite{2021MNRAS.504.1407R} has been carried out using the MeerKAT radio telescope array in South Africa \cite{2016mks..confE...1J, camilo_african_2018-1}. The results of the survey \cite{trapum_discoveries_url} included 13 MSPs in NGC~1851, a GC located in the Southern constellation of Columba\cite{2022A&A...664A..27R}. These include three massive, eccentric MSP binaries: PSR~J0514$-$4002A \cite{2004ApJ...606L..53F,2019MNRAS.490.3860R}, PSR~J0514$-$4002D and \E \cite{2022A&A...664A..27R}.

The latter has a spin period ($P$) of $5.6$ ms, an orbital period ($P_\mathrm{b}$) of $7.44$ days and an orbital eccentricity ($e$) of $0.71$ \cite{2022A&A...664A..27R}. The projected semi-major axis of the pulsar's orbit ($x\equiv a_\mathrm{p} \sin i/c$, where $a_\mathrm{p}$ is the semi-major axis of the pulsar orbit, $i$ the orbital inclination and $c$ is the speed of light in vacuum) is 27.8 seconds \cite{2022A&A...664A..27R}. The mass function ($f$) is thus:
\begin{equation}
f(m_\mathrm{p}, m_\mathrm{c}) 
\equiv \frac{{({m_\mathrm{c}}\:{\sin{i}})}^{3}}{(m_\mathrm{p} + m_\mathrm{c})^2} 
= 4{\pi}^2\,\frac{c^3}{G}\,\frac{x^3}{{P_\mathrm{b}}^2} 
= 0.41672 \pm 0.00022 \: \Msun, \label{eq:mass_function}
\end{equation}
where $m_\mathrm{c}$ is the mass of the companion, $m_\mathrm{p}$ is the mass of the pulsar and $G$ is the gravitational constant. The unit $\Msun$ is the mass of the Sun; we use the nominal solar mass adopted by the International Astronomical Union \cite{Prsa_2016}. All uncertainties are a confidence interval (C. I.) corresponding to a 68.3\% confidence level, unless otherwise stated. Assuming $m_\mathrm{p}\, \geq\, 1.17\: \Msun $ (corresponding to the lowest NS mass measured \cite{Martinez_2015} and a theoretical lower limit \cite{Suwa_2018}) and an edge-on orbit ($i = 90 ^{\circ} $), the mass function alone indicates $m_\mathrm{c} \geq 1.40\:\Msun$.

\section*{Radio timing observations}

\begin{figure}[!ht]
    \centering \includegraphics[width=\textwidth]{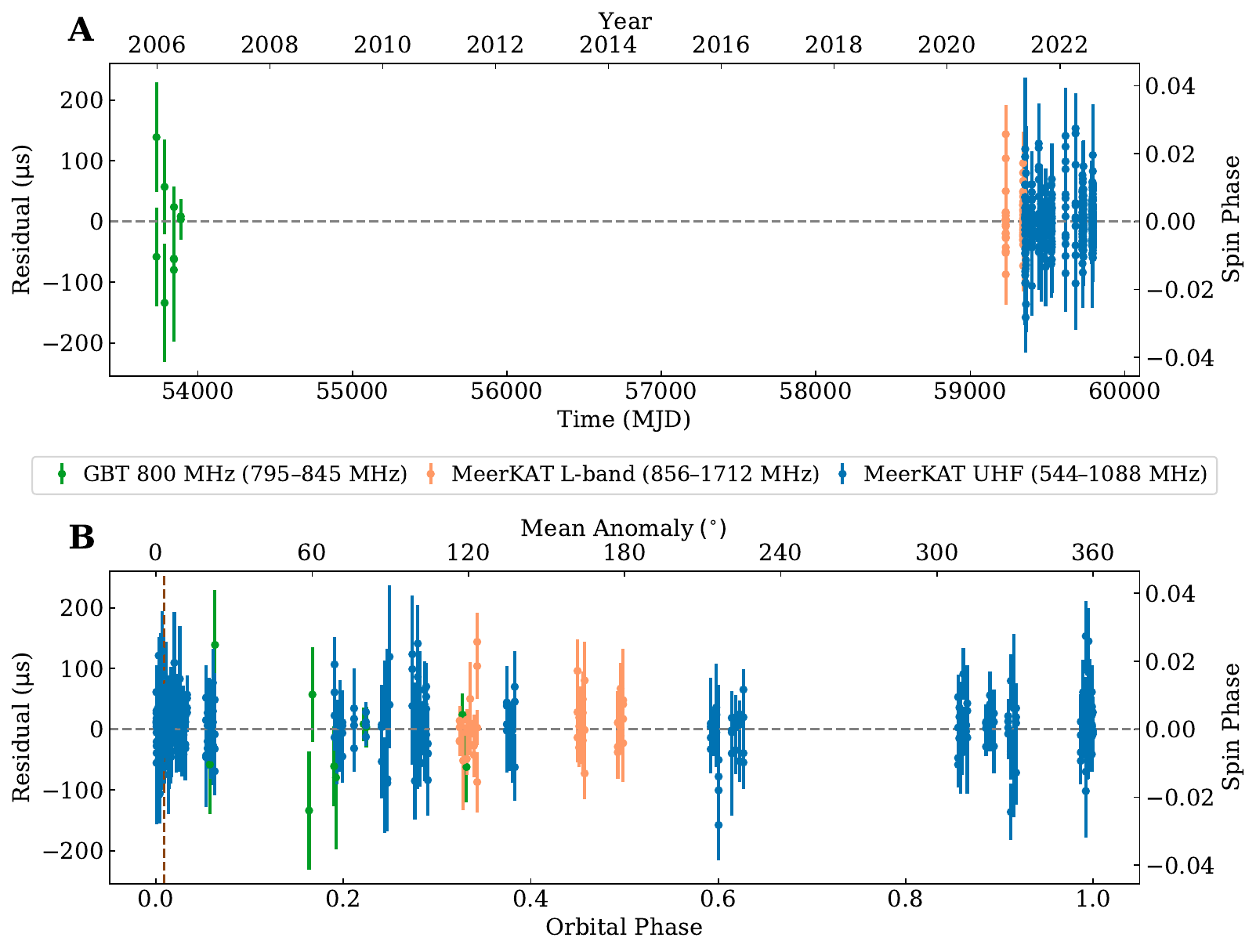}
    \caption{\textbf{Timing model fitting residuals for \E}. Shown are the residuals between the observed ToAs and the timing model presented in Table \ref{tab:ephemeris} as a function of observing epoch \textbf{(A)}, in units of Modified Julian Date (MJD), and orbital phase \textbf{(B)}. An orbital phase of 0 corresponds to periastron, and superior conjunction occurs at an orbital phase of 0.008 (shown with the brown dashed line). The vertical error bars indicate the 1-$\sigma$ uncertainties. The green points indicate data taken with the 800 MHz receiver on the GBT. Orange and blue points indicate data taken with the L-Band and UHF receivers of MeerKAT, respectively.}
    \label{Timing_Residuals}
\end{figure}

To determine the spin, astrometric and orbital parameters for all the pulsars in NGC 1851, we conducted 24 observations of this globular cluster using MeerKAT. The observations used either the L-band (856 to 1712 MHz) or Ultra High Frequency (UHF, 544 to 1088 MHz) receivers\cite{2021MNRAS.504.1407R} and were performed between January 2021 and August 2022. Data acquisition and initial reduction were performed using the Pulsar Timing User Supplied Equipment (PTUSE) instrument\cite{Bailes_2020}. We analysed the resulting times of arrival (ToAs) of the pulsed signal to determine an initial phase-coherent timing model for \E \cite{SOM}. Using this model, we recovered the previously undetected signals from this pulsar in six archival observations of NGC~1851 made with the 800 MHz (795 to 845 MHz) and S-band (1.73 to 2.60 GHz) receivers on the Robert C. Byrd Green Bank Telescope (GBT) \cite{2007ApJ...662.1177F} between December 2005 and August 2006. We combined the MeerKAT and GBT ToAs \cite{SOM} and re-fitted the timing parameters to determine a refined timing model. The results are listed in Table~\ref{tab:ephemeris} and the fitting residuals between this model and the observed ToAs are shown in Figure~\ref{Timing_Residuals}.

The timing model includes a precise measurement of the binary's rate of periastron advance, $\dot{\omega}\, = \,0.03468\, \pm \, 0.00003^{\circ}$  yr$^{-1}$. We obtain a consistent value (but slightly higher uncertainty) if we consider the MeerKAT data alone (Figure~\ref{fig:omegadot_variation}). As discussed in detail below, this effect is relativistic (with additional contributions being small relative to the measurement uncertainty (see Supplementary Text)) and its magnitude implies a high total system mass. We therefore performed an additional dense observing campaign \cite{SOM} to search for Shapiro delay, a relativistic light propagation delay in the system \cite{1964PhRvL..13..789S}. The longer time span provided by the GBT data also enabled us to search for the Einstein delay, an effect caused by the varying time dilation experienced by the pulsar at different orbital phases. We do not detect either the Shapiro delay or Einstein delay in our timing data, setting 95\% confidence upper limits of $h_3 < 1.48$ \textmu s and $\gamma_{\rm E} < 25$ ms, where $h_3$ is the orthometric amplitude of the Shapiro delay \cite{2010MNRAS.409..199F} and $\gamma_{\rm E}$ is the Einstein delay parameter. These non-detections further constrain the companion mass, as discussed below.

\begin{table}[]
   \centering
       \caption{\textbf{Timing model for {\E}.} The model was derived from the MeerKAT and GBT data with the single GBT observation at S-band excluded during fitting \cite{SOM}. Because the proper motion and parallax cannot be measured from the timing data, we adopt the bulk proper motion and parallax for NGC~1851 derived from HST and Gaia observations\cite{2022ApJ...934..150L}. Values are reported in the Dynamic Barycentric Time timescale and uncertainties are 68.3\% C.~I.}
   \begin{tabular}{ll}
       \toprule[0.1ex]
       \midrule[0.1ex]
       \multicolumn{2}{c}{Dataset and model fit quality} \\
       \midrule[0.1ex]
       Observation span\dotfill& MJD 53731 to 59793\\
       Number of ToAs\dotfill& 476\\
       Weighted root mean square residual\dotfill&  28.41 \textmu s\\
       Reduced $\chi^2$ value\dotfill& 1.019 \\
       Degrees of freedom\dotfill& 458\\
       \midrule[0.1ex] 
       \multicolumn{2}{c}{Fixed quantities}\\
       \midrule[0.1ex]
       Reference epoch\dotfill&  MJD 59400 \\
       Proper motion in right ascension, $\mu_{\alpha}$\dotfill& 2.128 mas yr$^{-1}$\\
       Proper motion in declination, $\mu_{\delta}$\dotfill& $-$0.646 mas yr$^{-1}$\\
       Parallax\dotfill& 0.0858 mas\\
       \midrule[0.1ex] 
       \multicolumn{2}{c}{Measured quantities}\\
       \midrule[0.1ex]
       Right ascension, $\alpha$ (J2000 equinox)\dotfill& 05$^{\rm h}$14$^{\rm m}$06$^{\rm s}$.73709 $\pm$ 0.00017 \\
       Declination, $\delta$ (J2000 equinox)\dotfill& $-$40$^{\circ}$02$'$48$''$.0556 $\pm$ 0.0014  \\
       Pulse frequency, $\nu$\dotfill& 178.70074989725 $\pm$ 0.000000000085 Hz\\
       First derivative of pulse frequency, $\dot{\nu}$\dotfill& (-6.1727 $\pm$ 0.0042) $\times$ 10$^{-15}$ Hz s$^{-1}$\\
       Second derivative of pulse frequency, $\ddot{\nu}$\dotfill& (7.3 $\pm$ 2.2) $\times$ 10$^{-26}$ Hz s$^{-2}$\\
       Dispersion measure, DM\dotfill& 51.93061 $\pm$ 0.00057 pc cm$^{-3}$\\
       Orbital period, $P_{\rm b}$\dotfill& 7.4478966582 $\pm$ 0.0000000072 days\\
       Projected semi-major axis of orbit, $x$\dotfill& 27.8192 $\pm$ 0.0050 s\\
       Orbital eccentricity, $e$\dotfill& 0.70793232 $\pm$ 0.00000085 \\
       Epoch of periastron, $T_0$ \dotfill& MJD 59361.29117138 $\pm$ 0.00000037 \\
       Longitude of periastron, $\omega_0$\dotfill& 65.317 $\pm$ 0.022$^{\circ}$ \\
       Rate of advance of periastron, $\dot{\omega}$\dotfill& 0.034676 $\pm$ 0.000031$^{\circ}$ yr$^{-1}$\\
       Rate of variation of the orbital period, $\dot{P}_\mathrm{b}$\dotfill& (18.1 $\pm$ 5.6) $\times$ 10$^{-12}$ s s$^{-1}$ \\
       Einstein delay, $\gamma_{\rm E}$\dotfill& 0.0111 $\pm$ 0.0084 s\\  
       Orthometric amplitude of Shapiro delay, $h_3$\dotfill& 0.02 $\pm$ 0.91 \textmu s\\
       \midrule[0.1ex] 
       \multicolumn{2}{c}{Derived quantities}\\
       \midrule[0.1ex]
       Total mass$^{*}$, $M$ \dotfill& 3.8870 $\pm$ 0.0045 $\Msun$\\ 
       Pulse period, $P$\dotfill&0.005595947418100 $\pm$ 0.000000000000027 s\\
       First derivative of pulse period, $\dot{P}$\dotfill& (1.9330 $\pm$ 0.0013) $\times$ 10$^{-19}$ s s$^{-1}$\\
       \bottomrule[0.1ex]
       \multicolumn{2}{l}{$^*$Assuming the observed $\dot{\omega}$ is due to relativistic effects (see Supplementary Text)}
    \end{tabular}
    \label{tab:ephemeris}    
\end{table}

\section*{Near-ultraviolet and optical observations}

If the $\geq 1.40 \, \Msun$ companion of \E\, were a main-sequence star, it should be detectable at optical wavelengths. We searched for an optical source using archival Hubble Space Telescope (HST) observations with the Wide Field Camera 3 (WFC3) in the F275W and F336W filters. This filter combination is particularly sensitive to blue stars such as blue straggler stars (BSSs) and WDs \cite{2020ApJ...895...15R,2021NatAs...5.1170C}. Bright BSSs are common in GCs, formed through stellar collisions or mass-transfer in a binary system; they can in principle have masses compatible with those predicted for
the companion of {\E}. 

\begin{figure}[!ht]
    \centering
    \includegraphics[scale=0.39]{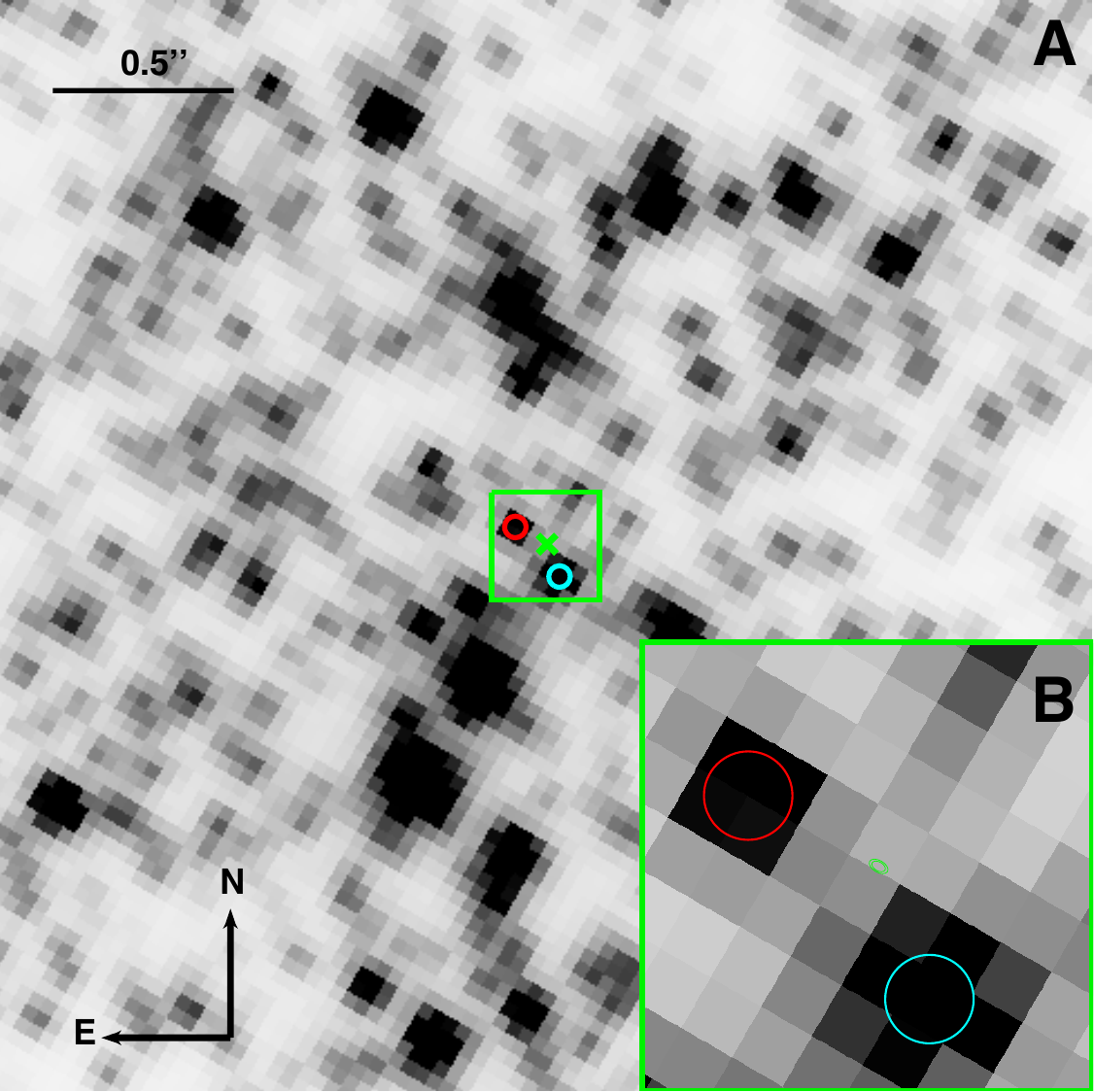}
    \includegraphics[scale=0.4]{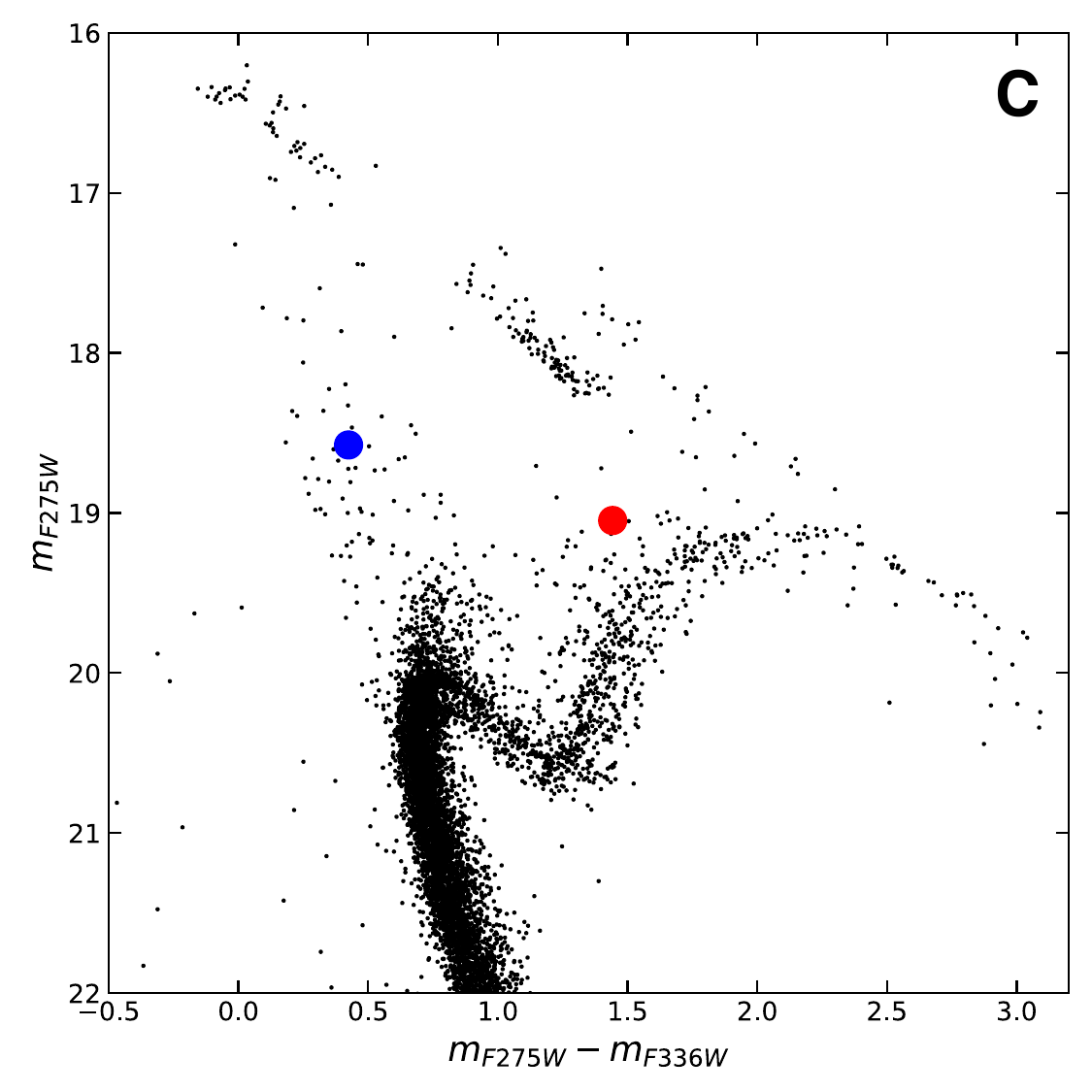}
    \caption{\textbf{Stars near the position of \E.} \textbf{(A)} HST image in the F275W filter of the $4'' \times 4''$ region surrounding the radio timing position of {\E} (green cross). The blue and red circles indicate two stars discussed in the text; the circle sizes are twice the astrometric uncertainty.
    \textbf{(B)} A $0.3'' \times 0.3''$ zoom of the green box of panel A.    
    The radio timing position is indicated with a green ellipse with size and orientation equal to  the 95\% C. I. on the pulsar's position relative to the ICRS.
    \textbf{(C)}  $m_{\mathrm{F275W}},m_{\mathrm{F275W}}-m_{\mathrm{F336W}}$ color-magnitude diagram of NGC~1851 derived from the HST images. The black dots represent all the cluster stars within the sampled field of view. The blue and red circles indicate the stars marked in panels A-B, which we interpret as a BSS and red giant, respectively. The photometric uncertainty for both the magnitude and color of the two marked stars is smaller than the symbol size \cite{SOM}.}
    \label{opticalobs}
\end{figure}

No optical source is detected at the position of {\E} (Figures~\ref{opticalobs}A and \ref{opticalobs}B). The closest stellar source  to the pulsar position is a star with a color-magnitude position consistent with those of BSSs (Figure \ref{opticalobs}C) and is offset by 90 milliarcseconds (mas), which is more than 6 times the astrometric precision (14 mas). It excludes a physical association, because the orbital separation between the pulsar and companion calculated from our timing model and located at the cluster distance of 11.66 kpc \cite{2022ApJ...934..150L}  is $<10^{-3}$ mas. Assuming the optical source is a hydrogen-burning star, we estimate its mass as $\sim 1.2\, \Msun$ \cite{SOM}, which is lower than the lower limit on $m_c$ from the mass function. Another star, located 100 mas from the pulsar position, is a red-giant. Red-giant branch stars in old GCs such as NGC 1851 have masses of about $0.7$--$0.8 \, \Msun$ \cite{Valcin_2020}, also lower than the minimum companion mass. We therefore conclude that the companion of \E~ is not detectable in the HST images.

\section*{System mass}
\label{sec:system_mass}

The non-detection of the companion of \E\ in the HST images implies that it must be a compact object. There is no measurable excess pulse dispersion near superior conjunction, which can be produced by ionized gas emanating from either a main sequence or a giant star companion\cite{SOM}. We therefore interpret the measured rate of advance of periastron ($\dot\omega$) as being purely of relativistic origin, with negligible contribution from the spins of the binary components (see Supplementary Text). Assuming general relativity (GR), the total mass of the system is then\cite{Robertson_1938,1982ApJ...253..908T}:
\begin{equation}
\label{eq:Mtot}
M \equiv m_\mathrm{p} + m_\mathrm{c} =
\frac{c^3}{G} \left[ \frac{\dot{\omega}}{3} (1 - e^2) \right]^{3/2}
\left( \frac{P_\mathrm{b}}{2 \pi}   \right)^{5/2} 
= 3.887 \:\pm\: 0.004 \: \Msun.
\end{equation}
This value is $1.0\, \rm M_{\odot}$ larger than the mass of the most massive double neutron star (DNS) known in the Milky Way, PSR~J1913+1102, which is $2.8887 \pm 0.0006 \, \Msun$ \cite{2020Natur.583..211F}. It is also larger than the total mass of the heaviest DNS merger detected in gravitational waves, GW190425, at $>99.5 \%$ probability [\cite{2020ApJ...892L...3A}, their figure 5].

\section*{Nature of the companion}

\begin{figure}[!ht]
    \centering
    \includegraphics[scale=0.55]{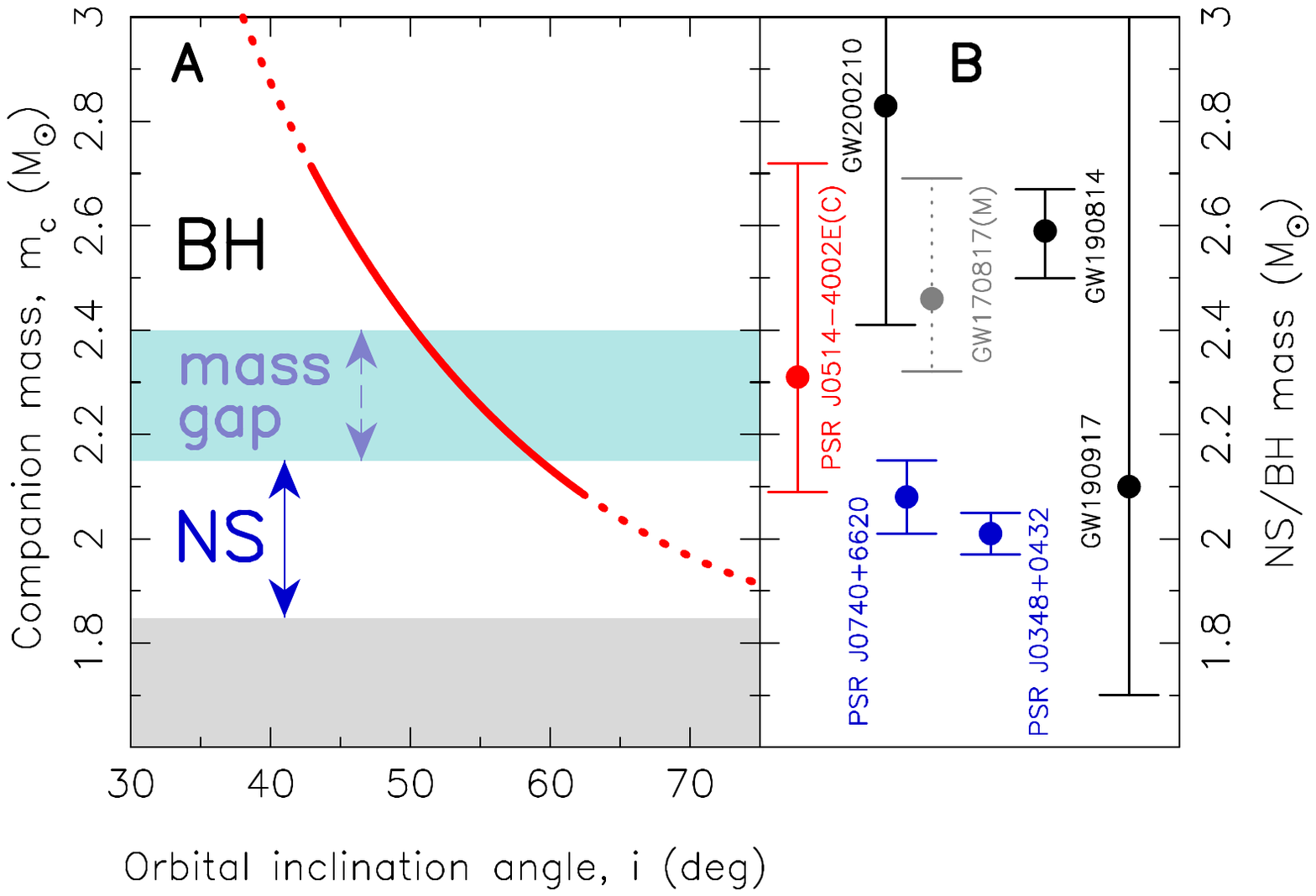}
\caption{{\bf Companion mass of \E.} {\bf (A)} Companion mass as a function of orbital inclination for the \E\ system. The solid red curve shows solutions within the 95\% C. I. (see text); the dotted part indicates lower and higher masses disregarded by our Bayesian model and assuming $m_p\ge 1.17\;M_\odot$, respectively. The grey-shaded region is ruled out because of the mass function and the total mass (Eqs.~\ref{eq:mass_function} and \ref{eq:Mtot}). Depending on the (unknown) NS equation-of-state \cite{Shibata_2019,Most_2020}, the light-blue shaded mass band corresponds to either massive NSs or light BHs. {\bf (B)} Inferred companion mass of \E\ (red) compared to the largest observed masses of radio pulsars (blue), low-mass components of gravitational-wave mergers (black), and the total post-merger remnant mass of GW170817 (grey, assuming no energy and mass loss after the inspiral, i.e.\ an absolute upper limit). Source names followed by (C) or (M) refer to the mass of the companion star and the remnant mass of the merger product, respectively; see Table~\ref{tab:mass-gap-references} for references, masses and error bars.}
    \label{fig:masses-NS-BH}
\end{figure}

Combining the measurement of the total mass with the mass function, we obtain, for an edge-on orbit ($i = 90^\circ$), $m_\mathrm{p}\, \leq \, 2.04 \: \Msun$ and $m_\mathrm{c} \geq 1.84\: \Msun$ (Figure ~\ref{fig:masses-NS-BH}). This companion mass is far too high for a WD, as the upper mass limit for a rigidly rotating WD is about $1.47 \, 
\Msun$ \cite{2005A&A...435..967Y}. Smaller inclination angles ($i < 90^\circ$) would imply a smaller $m_\mathrm{p}$ and larger $m_\mathrm{c}$. Adopting the minimum neutron star mass discussed above, $m_\mathrm{p} \geq 1.17\: \Msun $, we set limits of $i \geq 42.9^\circ$ and $m_\mathrm{c} \leq 2.71 \, \Msun$. 

We do not detect additional relativistic effects that could allow the individual masses to be determined. However, our upper limits on the Shapiro delay and the Einstein delay provide additional constraints on the masses and orbital inclination. To quantify these, we have made a Bayesian estimate of the component masses based on the quality of fit to the observational data ($\chi^2$) over a grid of total mass and orbital inclination values \cite{SOM}. The variation of the orbital period is contaminated by the acceleration of the system in the cluster. We assume that all other relativistic effects are as predicted by GR and that $m_\mathrm{p} \geq 1.17\: \Msun$ as above \cite{SOM}. We find Bayesian posteriors of $ M = 3.887 \:\pm\: 0.004 \: \Msun$, $m_\mathrm{p} = 1.53^{+0.18}_{-0.20} \, \Msun$, $m_\mathrm{c} = 2.35^{+0.20}_{-0.18}\, \Msun$ and $i = 52_{-5}^{+6 \, \circ}$ (median values with 68.3\% confidence limits \cite{SOM}). The 95\% probability limits are $i < 62^\circ$, $m_\mathrm{p} < 1.79\, \Msun$ and therefore $m_\mathrm{c} > 2.09 \, \Msun$. 
 
The companion mass is therefore likely to be in the mass gap for compact objects \cite{Littenberg_2015}, being higher than the largest precisely measured pulsar masses, $m_\mathrm{p} = 2.08\pm 0.07\: \Msun$ for PSR~J0740$+$6620 \cite{2021ApJ...915L..12F} and $m_\mathrm{p} = 2.01\pm 0.04\: \Msun$ for PSR~J0348$+$0432 \cite{2013Sci...340..448A}, while simultaneously below the observed minimum mass of about $5\;\Msun$ for BHs in Galactic X-ray binaries \cite{1998ApJ...499..367B,Ozel_2010}.

If the companion were a massive NS, it might also be a radio pulsar. We searched for radio pulsations from the companion, assuming the full allowed range of mass ratios, but did not detect any \cite{SOM}. We therefore cannot determine whether the companion is a massive NS or
a low-mass BH.

\section*{Formation of the system}

The combination of the location in a dense GC (where stellar exchange encounters often occur, see above) the highly eccentric orbit, the fast spin of the pulsar and the large companion mass indicates that the \E\ system is the product of a secondary exchange encounter. We propose that an earlier low-mass companion transferred mass to this pulsar, increasing the spin rate, before being replaced by the present high-mass companion in an exchange encounter. However, a more complicated evolution with multiple exchange encounters is also possible. We therefore cannot infer the nature of the companion from binary evolution models.

If the mass of the primary in \E\ is in the range 1.25 to 1.55 $\rm M_{\odot}$ of the four measurements of pulsar masses in GCs \cite{2006ApJ...644L.113J,2012ApJ...745..109L,2019MNRAS.490.3860R,2023A&A...671A..72C}, then the corresponding value of $m_\mathrm{c}$ (between $2.34$ and $2.63\, \rm M_{\odot} $) overlaps with the range of masses of remnants from mergers of DNSs, such as the merger product of GW170817 (Figure~\ref{fig:masses-NS-BH}). We suggest that, prior to becoming part of the current \E\ system, the companion could potentially have formed in such a merger event, regardless of whether it is a NS or BH. Although the probability of DNS mergers is generally low in GCs that have not undergone core collapse\cite{yfk+20}, NGC~1851 has a dense core (see Supplementary Text) which makes a DNS merger in the progenitor of the \E\ system more probable. A DNS has been observed in a GC (PSR B2127$+$11C, in M15) with a calculated merger time of 217 Myr \cite{2006ApJ...644L.113J}, implying that merger remnants are likely to be present in GCs.

Our derived companion mass overlaps with the mass estimates for the lighter components of the BH+BH or BH+NS merger candidates GW190814 \cite{2020ApJ...896L..44A}, GW190917 and GW200210 \cite{2021arXiv211103606T}. The lighter component of GW190814 has previously been interpreted as the product of an earlier merger \cite{2021MNRAS.500.1817L} that later acquired a more massive BH companion via
 exchange encounters in GCs (and then merged in the GW190814 event).

If the companion of \E\ is a light BH formed in such a merger, it would acquire a spin parameter $\chi_\mathrm{c} \in [0.6, 0.875]$ during the merger \cite{Bernuzzi_2020}, where $\chi_\mathrm{c}$ is the dimensionless BH spin angular momentum. A NS rotating at the maximum theoretical rate would have a similar $\chi_\mathrm{c}$ immediately after merger \cite{Lo_2011}, though we expect this would decrease rapidly after formation due to electromagnetic torque. Assuming a magnetic field of $10^9\, \rm G$, the spin parameter would become $\lesssim 0.3$ (corresponding to the fastest known MSPs) after $\sim 30\, \rm Myr$, so we regard a fast-spinning NS companion as unlikely.

A BH companion with $\chi_\mathrm{c} \in [0.6, 0.875]$ would induce relativistic spin-orbit coupling, causing the orbital plane to precess around the total angular momentum vector, an effect known as Lense-Thirring precession. We calculate the resulting variation of the projected semi-major axis of the pulsar's orbit ($\dot{x}$) would be  $\lesssim 1.7 \times 10^{-13}$. This is slightly beyond the effect size that would be detectable in our data; the 1-$\sigma$ uncertainty on the measurement of $\dot{x}$ is $2.0 \times 10^{-13}$ (see Supplementary Text). We therefore cannot differentiate between a NS and a BH companion.

\bibliographystyle{Science}
\bibliography{scibib}

\section*{Acknowledgments}

\paragraph{Acknowledgments} 

The MeerKAT telescope is operated by the South African Radio Astronomy Observatory (SARAO), which is a facility of the National Research Foundation, an agency of the Department of Science and Innovation. SARAO acknowledges the ongoing advice and calibration of GPS systems by the National Metrology Institute of South Africa (NMISA) and the time space reference systems department of the Paris Observatory. Parts of this research were conducted by the Australian Research Council Centre of Excellence for Gravitational Wave Discovery (OzGrav), through project number CE170100004. MeerTime data is housed on the OzSTAR supercomputer at Swinburne University of Technology. The OzSTAR program receives funding in part from the Astronomy National Collaborative Research Infrastructure Strategy (NCRIS) allocation provided by the Australian Government. PTUSE was developed with support from the Australian SKA Office and Swinburne University of Technology. The authors also acknowledge Max-Planck-Institut f\"{u}r Radioastronomie funding to contribute to MeerTime infrastructure. The National Radio Astronomy Observatory is a facility of the National Science Foundation operated under cooperative agreement by Associated Universities, Inc. The Green Bank Observatory is a facility of the National Science Foundation operated under cooperative agreement by Associated Universities, Inc. TRAPUM observations use the FBFUSE and APSUSE computing clusters for data acquisition, storage and analysis. These clusters were funded, installed and operated by the Max-Planck-Institut f\"{u}r Radioastronomie and the Max-Planck-Gesellschaft. This work has made use of data from the European Space Agency (ESA) mission {\it Gaia} (\url{https://www.cosmos.esa.int/gaia}), processed by the {\it Gaia} Data Processing and Analysis Consortium (DPAC, \url{https://www.cosmos.esa.int/web/gaia/dpac/consortium}). Funding for the DPAC has been provided by national institutions, in particular the institutions participating in the {\it Gaia} Multilateral Agreement.

\paragraph{Funding}

AC, AP, AR, and MBu acknowledge the use of resources from the research grant "iPeska" (P.I. Andrea Possenti) funded under the INAF national call Prin-SKA/CTA approved with the Presidential Decree 70/2016. AP and AR also had the support from the Ministero degli Affari Esteri e della Cooperazione Internazionale - Direzione Generale per la Promozione del Sistema Paese - Progetto di Grande Rilevanza ZA18GR02. SMR is a CIFAR Fellow and is supported by NSF Physics Frontiers Center awards 1430284 and 2020265. EDB, AD, PCCF, MK, VVK, NW, JB, WC, DJC, C.-H.R.C., YPM and PVP acknowledge continuing valuable support from the Max-Planck Society. BWS acknowledges funding from the European Research Council (ERC) under the European Union's Horizon 2020 research and innovation programme (grant agreement No. 694745). C.-H.R.C. acknowledges support from the Deutsches Zentrum f\"{u}r Luft- und Raumfahrt (DLR) grant NS1 under contract no. 50 OR 2214. C.P. and M.C. acknowledge funding from Italian MIUR throughout the PRIN-2017 grant awarded to the Light-on-Dark project (PI: Ferraro) through contract PRIN-2017K7REXT. MG acknowledges funding from the South African Research Chairs Initiative of the Department of Science and Technology and the National Research Foundation of South Africa.

\paragraph{Author contributions}

Conceptualization, AD, AP, BWS, MB, MK, NW, PCCF, SMR, VVK; Data curation, AD, AR, EDB, MB, MG, PVP, SMR, YPM, VVK; Formal Analysis, AD, AR, C-HRC, CP, DJC, MBu, MC, NW, PCCF, TG, VVK; Funding acquisition, AP, MB, MK; Investigation, AD, AR, C-HRC, EDB, MBu, SB, VVK; Methodology, AD, EDB, MB, PCCF, SMR; Project administration, AD, AP, AR, BWS, EDB, MB, MBu, MK, PCCF; Resources, AR, BWS, EDB, JB, MB, MBu, MK, PCCF, C-HRC; Software, AR, DJC, EDB, MB, MG, PCCF, PVP, SB, WC, SMR; Supervision, AP, BWS, DJC, EDB, MK, NW, PCCF, SMR, VVK; Validation, AD, C-HRC, NW, PCCF, TMT, VVK; Visualization, AD, EDB, MC, PCCF, TMT, VVK; Writing – original draft, AD, AR, C-HRC, CP, EDB, MC, NW, PCCF, TMT, VVK; Writing – review \& editing, AC, AD, AP, BWS, C-HRC, CP, DJC, EDB, MB, MC, MG, MK, NW, PCCF, PVP, SMR, TG, TMT, VVK.

\paragraph{Competing interests}

The authors have no competing interests to declare.

\paragraph{Data and materials availability} 

Our measured ToAs, timing model, and simulation code are available at https://github.com/ewanbarr/NGC1851E\_Additional\_Materials/ and archived at Zenodo \cite{github_repo}. Folded pulsar timing archives from MeerKAT and the GBT as well as dedispersed pulsar search mode observations from MeerKAT are available at https://keeper.mpdl.mpg.de/d/a59d968eccf148239f1d/ and archived at the Max Planck Digital Library \cite{keeper_repo}. Optical and near-ultraviolet data sets can be downloaded from the MAST archive at https://mast.stsci.edu/search/ui/\#/hst, using the search terms ``Target=NGC 1851'', ``Proposal ID=12311,13297'' and ``Filters / Gratings=F275W,F336W''.

\section*{Supplementary Material}
Materials and Methods\\
Supplementary Text\\
Figs. S1 to S10\\
Tables S1 and S2\\
References (51–108)

\cleardoublepage

\setcounter{figure}{0}
\makeatletter 
\renewcommand{\thefigure}{S\@arabic\c@figure}
\makeatother

\setcounter{table}{0}
\makeatletter 
\renewcommand{\thetable}{S\@arabic\c@table}
\makeatother

\setcounter{equation}{0}
\makeatletter 
\renewcommand{\theequation}{S\@arabic\c@equation}
\makeatother


\setcounter{page}{1}


\section*{Materials and Methods}

\subsection*{Radio timing observations and data analyses}
\subsubsection*{MeerKAT observations}

\E\ was detected with the MeerKAT radio telescope on 24 occasions between 2021 and 2022, and all observations were conducted using the standard configuration for globular cluster observations \cite{2021MNRAS.504.1407R,Abbate_2022}. The initial three MeerKAT observations were conducted with the L-band receivers, covering the band 856 to 1712 MHz, and the following 21 observations using the UHF receivers covering the band 544 to 1088 MHz. Data acquisition used the search mode of the Pulsar Timing User Supplied Equipment (PTUSE) instrument \cite{Bailes_2020}. This mode recorded time and frequency resolved data that was coherently dedispersed \cite{2011PASA...28....1V} to the nominal dispersion measure of the cluster, 52.14 pc cm$^{-3}$ (the DM of the only pulsar previously known in the cluster, \cite{2019MNRAS.490.3860R}). Depending on the array configuration, MeerKAT data have time resolutions between 7 and 15 \textmu s. Observations varied in duration between 2 and 4 hours.

To maximise sensitivity to Shapiro delay, six of the UHF observations were scheduled to coincide with the orbital phases of the expected extrema of the unabsorbed Shapiro delay signal (the part of the Shapiro delay that is not easily absorbed by varying the Keplerian orbital parameters [\cite{2010MNRAS.409..199F}, especially their eq. 31] within a single orbit). These phases were determined by synthesising low-noise ToAs using the timing model in Table~\ref{tab:ephemeris} with the inclusion of a detectable Shapiro delay signal. The ToAs were then fit for the Keplerian orbital parameters with a timing model which assumed no Shapiro delay. The extrema of the unabsorbed Shapiro delay signal are then taken as the extrema of the model residuals from the fit. Synthesising the ToAs using Shapiro delay signals from a range of possible geometries showed negligible variation in the orbital phases corresponding to the extrema. These observations are highlighted in Table~\ref{tab:observations}.

\subsubsection*{GBT observations}

Using a preliminary timing model for \E\ derived from only MeerKAT observations (see below), we could recover the signal of the pulsar with the GBT on 6 occasions between 2005 and 2006. Of these detections, 5 were made in observations conducted with the 800 MHz receiver covering the band 795 to 845 MHz and one in an observation conducted with the S-band receiver covering the band 1649.6 to 2249.6 MHz. For both receivers the Pulsar Spigot instrument was used for data acquisition \cite{2005PASP..117..643K}. Data were recorded with a time resolution of 82 \textmu s. Observations varied between 80 minutes and 4 hours in duration. Table~\ref{tab:observations} summarizes all the GBT and MeerKAT observations used in this work.

\subsubsection*{Data reduction}

\begin{figure}[H]
    \centering \includegraphics[width=\textwidth]{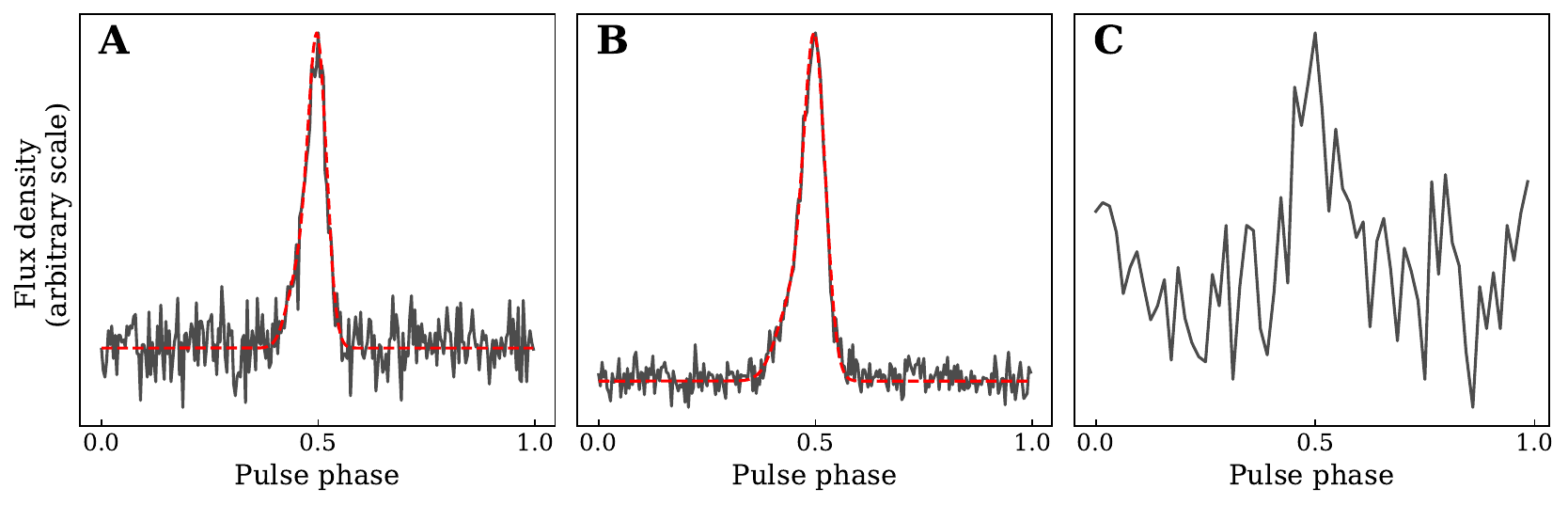}
    \caption{\textbf{Multi-frequency pulse profiles for \E\ }. The summed profiles are shown in grey, and the red dashed lines show the corresponding analytic profiles. \textbf{(A)} Pulse profile for MeerKAT L-Band (856 to 1712 MHz) generated by summing the profiles for all observations. \textbf{(B)} Same as (A), but for MeerKAT UHF (544 to 1088 MHz). \textbf{(C)} Pulse profile of the highest S/N observation for GBT 800 MHz (795 to 845 MHz).}
    \label{Profile_Plots}
\end{figure}

The coherently dedispersed MeerKAT search-mode data were folded with 256 bins across the pulse profile using the \textsc{dspsr} software \cite{2011PASA...28....1V, dspsr_url} with the timing ephemeris of \E\, from \cite{2022A&A...664A..27R}. These folded archives were then downsampled to 256 frequency channels and manually cleaned for any residual radio frequency interference (RFI) using the routines \texttt{psrzap} 
and \texttt{pazi}, part of the \textsc{psrchive} software package \cite{Hotan+2004, psrchive_url}. For both the L-band and UHF observations, a standard profile template was generated by summing the pulse profiles from the corresponding observations to generate a low-noise profile. An analytic profile was then generated from this high signal-to-noise ratio (S/N) profile for each observing band by fitting the low-noise profile with two components, modelled by von Mises functions, using the \texttt{paas} program from \textsc{psrchive}. The integrated pulse profile and analytic profile for the L-Band and UHF are shown in Figures~\ref{Profile_Plots}A and \ref{Profile_Plots}B, respectively. Although we use separate templates for UHF and L-band, no profile difference is observed between the bands.

The pulse profiles for the individual observations were then cross-correlated with the corresponding analytic profiles for each band, and ToAs were extracted using the \texttt{pat} command in \textsc{psrchive}. The ToAs were derived with 4 sub-bands and integrations of 10 to 60 minutes, depending on the length of the observation and S/N of the detection.

Data acquired with the GBT were incoherently dedispersed and phase folded with 64 bins across the profile using the \texttt{prepfold} routine of the \textsc{presto} software package \cite{Presto, presto_url} using the ephemeris of \E\, derived from the MeerKAT observations. ToA estimation used a previously described method \cite{2007ApJ...662.1177F}. For ToA estimation, a standard profile template was generated from the highest S/N GBT observation with the 800 MHz receiver (Figure~\ref{Profile_Plots}C). Due to the unknown profile evolution between 820 MHz and 1950 MHz, only data from the 800 MHz receiver were used in the timing analyses.

\subsubsection*{Phase coherent solution}

We used the \textsc{tempo}\cite{tempo_url} timing software for the analysis of the ToAs. 
Initially, we extended the previous orbital solution \cite{2022A&A...664A..27R} to all available MeerKAT ToAs. At this stage, we fitted arbitrary time offsets between each group of ToAs from each observation, because the rotation count between observations was not known due to insufficient precision in the previously-determined timing parameters. 

In a second stage, we removed these arbitrary offsets using the \texttt{DRACULA} algorithm \cite{2018MNRAS.476.4794F}. This allowed us to determine the rotation counts between observing epochs and, hence, a phase-coherent timing solution for all the MeerKAT data.

Using this preliminary solution, we re-folded the archival data from the GBT observations \cite{2007ApJ...662.1177F}. We detect the pulsar in six GBT observations taken with the 800 MHz and S-band receivers (Table~\ref{tab:observations}). These detections are faint, so we only determine two ToAs from each detection. In the subsequent analysis, we have used only the ToAs from the five observations taken with the 800-MHz receiver.

Given the $>15$-year gap between the last GBT and first MeerKAT detections of the pulsar, we investigate the validity of the MeerKAT-determined rotational model for the GBT observing epochs. To test this, we used \texttt{DRACULA} to determine the combination of missing rotations between the GBT observations that minimises the $\chi^2$ of the ToA residuals. The algorithm returned 32 possible solutions with reduced $\chi^2 < 2$. Of these, the most favoured model is that of no missing rotations over the GBT epochs, which has reduced $\chi^2 = 1.019$. The next best solution consists of a single phase jump with reduced $\chi^2 = 1.255$. For the 458 degrees of freedom in the model, these correspond to $\chi^2 = 466.70$ and $\chi^2 = 561.05$, respectively. Using equation~\ref{eq:likelihood} below, we find that the next best solution has $3.2 \times 10^{-21}$ times the Bayesian likelihood of the solution in which the GBT observations are phase connected.  

For all the datasets, we used pulse profile templates that have, as much as possible, consistent phase definitions. The profiles used for the GBT and the MeerKAT data were aligned by eye to a phase precision of 0.01 to ensure that no large arbitrary time offsets are introduced by \textsc{tempo} due to different physical longitudes on the neutron star to which the ToAs refer. \textsc{tempo} assumes that all ToAs refer to a consistent longitude; if there are differences in the arrival times caused by different phase definitions, the software assumes that there is a time delay between the systems, something that can bias other parameters, like $T_0$ \cite{Guo_2021}.

This consistent phase definition allows us to determine the offset between the GBT and the MeerKAT ToAs, which is $0.6 \pm 1.3\,$ms. This is consistent with zero and smaller than the spin period, consistent with phase connection between the two data sets. The parameters in Table~\ref{tab:ephemeris} are derived by fitting for this time offset parameter. This results in slightly larger uncertainties than in the case where we do not fit for this offset parameter, therefore a more conservative estimate of parameter precision. This small offset suggests that, within our measurement precision, both observing systems share functionally identical clocks. If we assume that the offset is zero, then the number of rotations between the last 2006 ToA and the first 2021 ToA is 82,461,896,322; this provides a $\chi^2$ of 473, subtracting or adding 1 rotation we obtain $\chi^2 = 484$. Using equation~\ref{eq:likelihood} below, we obtain for each of the two latter solutions 0.0041 times the Bayesian likelihood of the former solution; this implies the former has an overall probability of 99.19\%. Additional solutions have negligible probabilities. This number of rotations corresponds to 461,453,260.433689 s, or 14.62 years. The estimated timing parameters under this assumption are consistent within 1-$\sigma$ with the parameters in Table~\ref{tab:ephemeris}. However, since we cannot rule out time offsets smaller than about 1.3 ms, we regard the parameters in Table~\ref{tab:ephemeris} as more reliable, since they already take the uncertainty of this small offset into account. 

We also investigated including a systematic time offset between MeerKAT L-band and UHF observations. The best-fitting value of $18\pm9$ \textmu $\mathrm{s}$ is consistent with an offset arising from the mis-alignment of the profile templates between the two bands. The instrumental offset between the two bands is expected to be less than 5 ns \cite{Bailes_2020}.

\subsubsection*{Timing parameters}

We used \textsc{tempo} to analyze the ToAs, which were first converted to the Bureau International des Poids et M\'esures (BIPM) 2019 terrestrial time standard (BIPM2019). Then, we use the Jet Propulsion Laboratory's DE440 Solar System ephemeris \cite{2021AJ....161..105P} to compensate for the motion of the radio telescopes relative to the Solar System barycentre (SSB). The results are reported in the Dynamic Barycentric Time (TDB) timescale. To describe the effects of the orbital motion and light propagation, we use a modification of the theory-independent DD orbital model \cite{DD_1986,Damour_Taylor_92} that includes the orthometric parametrization of the Shapiro delay \cite{2010MNRAS.409..199F} (hereon the DDH model). Like the DD model, the DDH model can describe the timing of binary pulsars for a wide range of alternative theories of gravity \cite{Damour_Taylor_92}.
Table \ref{tab:ephemeris} lists the resulting timing parameters. The ToA residuals from the best fitting model as a function of observing epoch and orbital phase are shown in Figure~\ref{Timing_Residuals}. No trends are apparent, and the residuals are consistent with the ToA uncertainties. The overall best fitting model has a reduced $\chi^2$ of 1.019, which was obtained after adjusting the uncertainties of individual data sets such that the reduced $\chi^2$ for each data set is 1.0. The adjustment factors for each dataset were: GBT 800 MHz data: 1.227 (indicating that the original uncertainties are slightly under-estimated), MeerKAT L-band data: 0.989, MeerKAT UHF data: 1.012. These adjustments result in more conservative estimates of parameter precision. They are close to unity, implying the ToA uncertainties were well estimated by the routines used to derive the ToAs. The overall reduced $\chi^2$ is close to 1, indicating that the timing solution provides a statistically justified description of the observed ToAs.

We cannot measure the proper motion of \E\, because the MeerKAT data only span a period of 1.5 years. We therefore assume that the pulsar's proper motion matches that of NGC~1851 determined from an analysis of HST and Gaia astrometric data  \cite{2022ApJ...934..150L}: $\mu_{\alpha} = 2.128 \pm 0.031 \, \rm mas \, yr^{-1}$, $\mu_{\delta} = -0.646 \pm 0.032 \, \rm mas \, yr^{-1}$. We expect that the proper motion of \E\ might differ slightly from this value, but that difference should be small, given the fact that the GC has a central escape velocity of only $42.9\:\rm{km}\: \rm{s}^{-1}$ \cite{BaumgardtHilker18}, corresponding to a proper motion of 0.78 mas yr$^{-1}$. We also assume a parallax of $0.0858\,\pm\,0.0018 \, \rm mas$, which corresponds to the astrometric distance estimate of $11.66\pm0.25$~kpc for NGC 1851 \cite{2022ApJ...934..150L}.

The timing model includes a 3-$\sigma$ detection of the second derivative of the pulsar's spin frequency, $\ddot{\nu} =  (7.3 \pm 2.2) \times 10^{-26}$ Hz s$^{-2}$. This is likely to be caused by the variation of the acceleration of the system, caused by its change of position relative to the gravitational potential of the GC, or relative to nearby stars \cite{1992RSPTA.341...39P}. Such changes have been seen for other GC pulsars with long-term timing, and are consistent with theoretical expectations  \cite{2017MNRAS.471..857F,2017ApJ...845..148P,2018MNRAS.476.4794F}. Our measurement of $\ddot{\nu}$ is of a similar magnitude to the values observed for pulsars in other GCs \cite{2017MNRAS.471..857F,2017ApJ...845..148P}. Introducing this parameter results in slightly increased uncertainties of all other timing parameters, especially those that improve more with the use of the early GBT data: $\dot{\nu}$, $\dot{P}_{\rm b}$ and $\gamma_{E}$.

In the orbital model we adopted, relativistic effects detectable in the timing are quantified by post-Keplerian (PK) parameters. We measure $\dot{\omega}$ with $>$ 1000-$\sigma$ significance even if we use only the MeerKAT data, and obtain an upper limit for the orthometric amplitude of the Shapiro delay, $h_3$. We fit the DDH model including both Shapiro delay parameters $h_3$ and $\varsigma$ (the orthometric ratio of the Shapiro delay), but find the model does not converge because the Shapiro delay is not detected. We therefore fixed one of the parameters, $\varsigma$, to a value of 0.456, which corresponds to an inclination of $49^\circ$ from our ephemeris (see below). The assumed value of $\varsigma$ has little effect on the detection of $h_3$ because the two parameters are designed to be only weakly correlated \cite{2010MNRAS.409..199F}. Using this fixed $\varsigma$, we obtain $h_3 = -0.02 \pm 0.91$ \textmu s, which is a non-detection with a 2-$\sigma$ upper limit of $1.80$ \textmu s.

Including the earlier GBT data provides a 3-$\sigma$ detection of the variation of the orbital period, $\dot{P}_\mathrm{b}\, = \, (18.1 \, \pm\, 5.6)\, \times 10^{-12}$ s s$^{-1}$ and a 2-$\sigma$ upper limit for the Einstein delay, $\gamma_{\rm E} < 28\, \rm ms$. The latter parameter quantifies the variation of a combination of the special relativistic time dilation and gravitational redshift with orbital phase.

The $\dot{\omega}$ parameter quantifies a relativistic effect which can be used to estimate the total system mass, $M$, if we assume GR is the correct theory of gravity. However, this is not the case for $\dot{P}_\mathrm{b}$. For some binary pulsars, including PSR~B1913+16 and PSR~J0737$-$3039A, the observed $\dot{P}_\mathrm{b}$ is dominated by gravitational wave damping \cite{2016ApJ...829...55W,2021PhRvX..11d1050K}. However, for wider binaries, this effect becomes insignificant; in the case of \E\, the GR prediction for $\dot{P}_\mathrm{b}$ corresponding to the most likely masses is $\dot{P}_{b,\, \rm GR}\, = \, -0.05 \, \times \, 10^{-12}$ s s$^{-1}$, hundreds of times smaller than the observed $\dot{P}_\mathrm{b}$. The variation of $P_\mathrm{b}$ is instead likely to be dominated by the secular variation of the Doppler shift: this does not include the orbital variation of the Doppler shift, which is already subtracted by the timing model. This secular variation of the Doppler shift is itself dominated, in this case, by the line-of-sight component of the acceleration of the system in the gravitational field of the GC, $A$. Assuming this is the case, 
\begin{equation}
\label{eq:a1}
A \simeq \frac{\dot{P}_\mathrm{b}}{P_\mathrm{b}} c = (8.4 \pm 2.6) \times 10^{-9}\, \rm m \, s^{-2}.
\end{equation}
This acceleration also affects the variation of the spin period derivative $\dot{P}$. Our measured value (Table~\ref{tab:ephemeris}) implies a characteristic age of 0.46~Gyr, which is within the range of previously studied MSPs \cite{2023pbse.book.....T} but unusually low. If, instead, $\dot{P}$ is dominated by $A$, then
\begin{equation}
 \label{eq:a2}
A \simeq \frac{\dot{P}}{P} c = 10.3 \times 10^{-9}\, \rm m \, s^{-2},
\end{equation}
which is consistent with the estimate of $A$ above. We subtract the estimate of $A$ from equation~\ref{eq:a1} from the observed spin-down to estimate the intrinsic spin-down of the pulsar ($\dot{P}_{\rm int}$), independently from the secular variation of the Doppler shift:
\begin{equation}
    \dot{P}_{\rm int} = \dot{P} - \frac{\dot{P}_\mathrm{b}}{P_\mathrm{b}} P = (3.6 \pm 4.8) \times 10^{-20}\, \rm s\, s^{-1},
\end{equation}
which is a non-detection with 2-$\sigma$ upper limit of $13.2\times10^{-20}$~s~s$^{-1}$. Thus, the intrinsic spin-down of \E\ is too small to be measured from our data. 

\subsubsection*{Robustness of the timing parameters}

\begin{figure}[H] 
  \includegraphics[width=\textwidth]{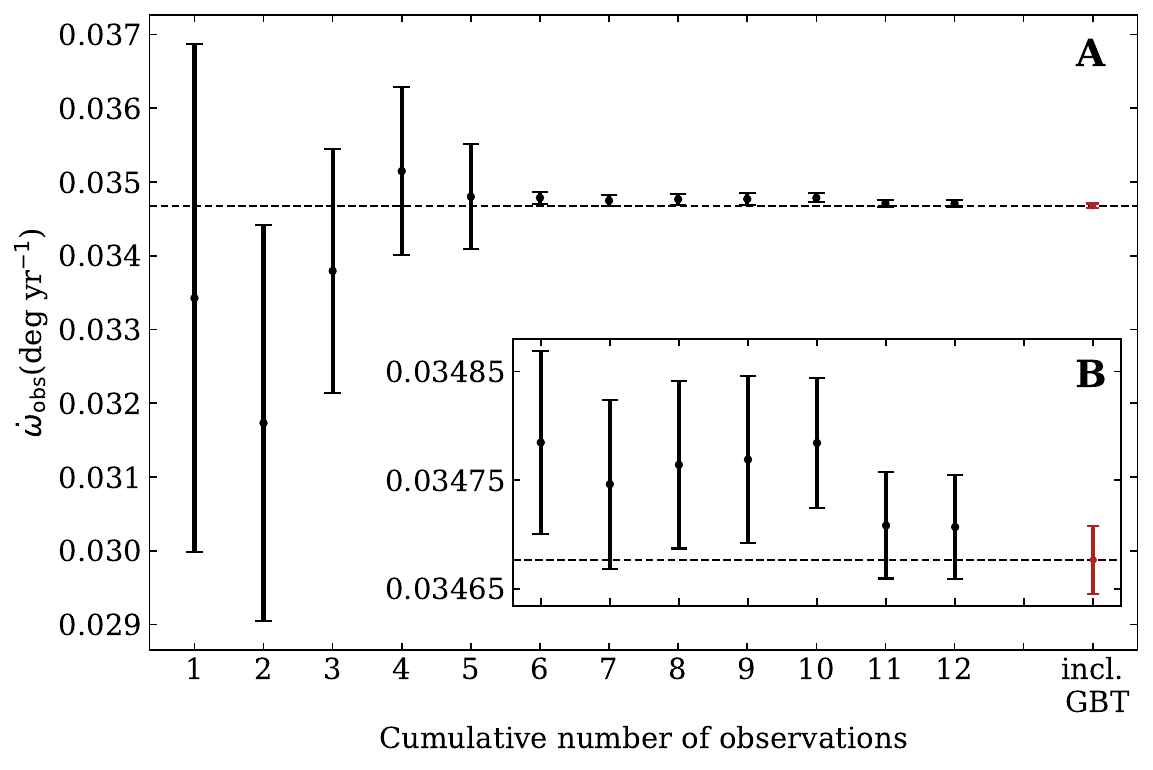}
  \caption{\textbf{Observed rate of advance of periastron ($\dot\omega_{\mathrm{obs}}$) for \E.} In both panels, the vertical errorbars indicate 1-$\sigma$ uncertainties. \textbf{(A)} $\dot\omega_{\mathrm{obs}}$ as a function of the cumulative addition of observations from the MeerKAT dataset. \textbf{(B)} Same as panel (A), but only for the last 7 observations from the MeerKAT dataset. The last point indicates the value of $\dot\omega$ derived with the addition of the GBT data set to the full MeerKAT dataset, and the horizontal line corresponds to this value as reported in Table \ref{tab:ephemeris}.
  } 
  \label{fig:omegadot_variation}
\end{figure}

We investigate the robustness of our measurement of the rate of advance of periastron of the pulsar by performing a self-consistent check of the observed values as a function of the number of observations. The results are shown in Figure~\ref{fig:omegadot_variation}; we find no substantial change in $\dot{\omega}$ as more observations are added and the fluctuations are consistent with the uncertainties obtained at each stage. This stability indicates that the value of $\dot{\omega}$ is robust.

\begin{figure}[H] 
  \centering \includegraphics[scale=0.8]{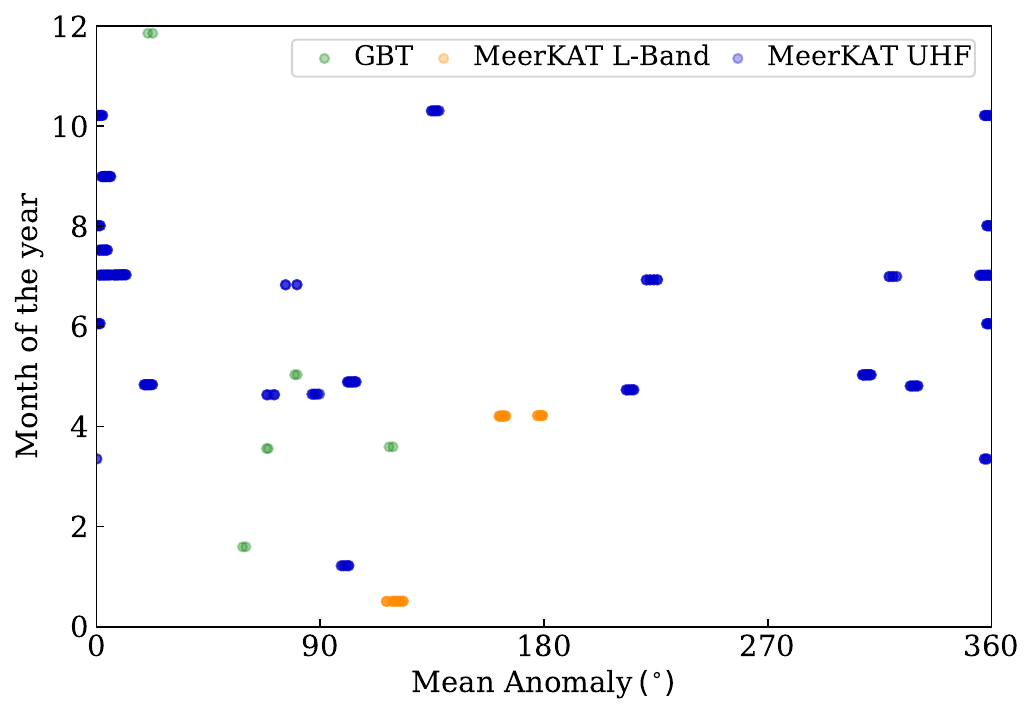}
  \caption{\textbf{Orbital and monthly coverage of \E\, observations.} Each point shows a single ToA estimation as a function of the time of year it was obtained and the mean anomaly of \E's orbit to which it corresponds. A mean anomaly of 0 denotes periastron passage. Green points indicate ToAs estimated from the archival GBT dataset, while the orange and blue points indicate ToAs estimated from the MeerKAT L-Band and the UHF datasets, respectively.} 
  \label{fig:Months_OrbitalPhases}
  \end{figure}

As an additional check, we investigated the orbital coverage of all the observations (Figure~\ref{fig:Months_OrbitalPhases}). We find a broad distribution of the observations in orbital phase and time of the year, which improves the determination of $\dot{\omega}$ and the position and reduces the correlations between orbital and astrometric quantities. 

\begin{figure}[H] 
  \centering \includegraphics[width=\textwidth]
  {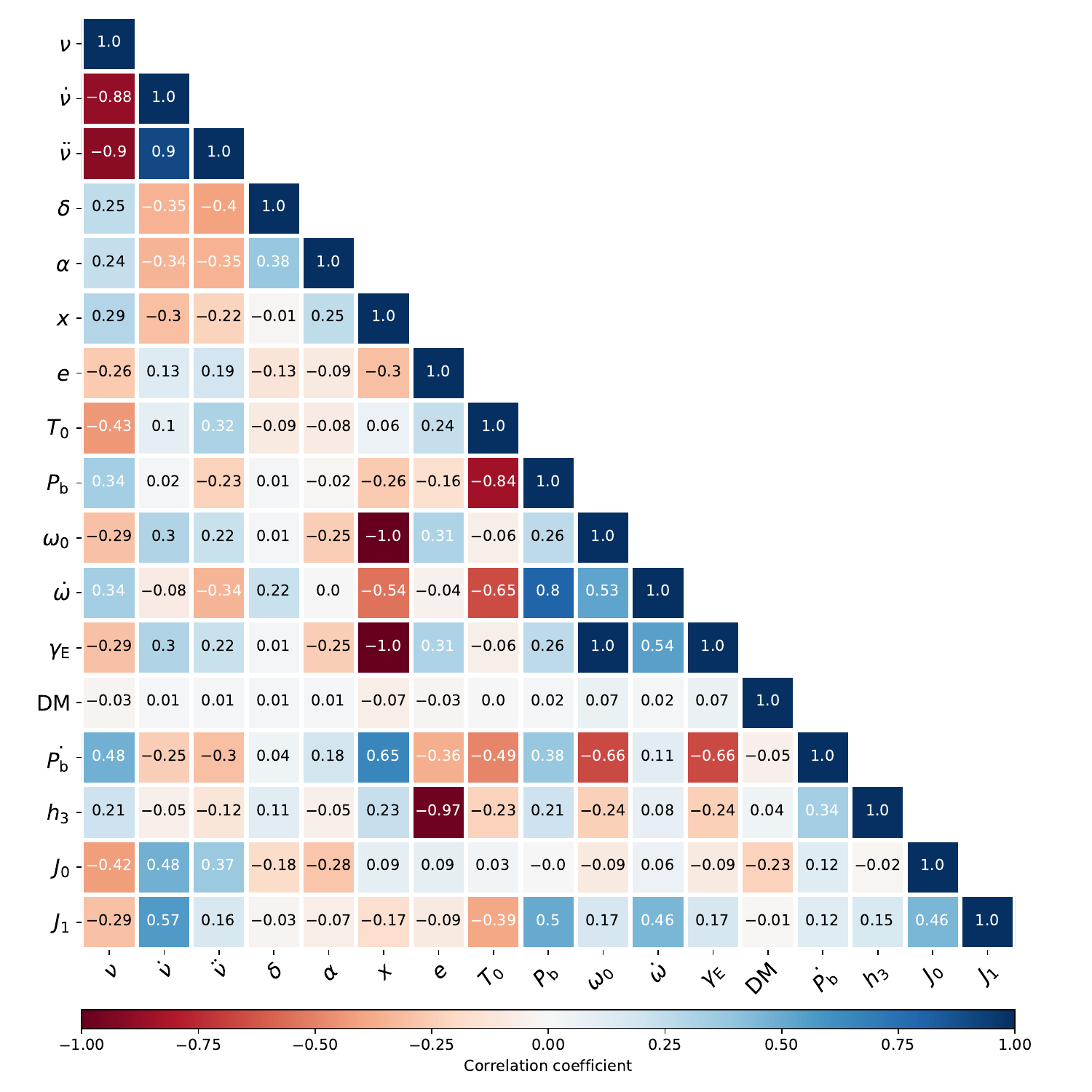}
  \caption{\textbf{Correlations between the timing parameters of \E.} Here we show the Pearson product-moment correlation coefficient (colour scale) between the timing parameters presented in Table~\ref{tab:ephemeris}, as reported by {\sc tempo}. We also include $J_0$, the time offset between MeerKAT L-band and MeerKAT UHF data, and  $J_1$, the time offset between GBT data and MeerKAT UHF data.} 
  \label{fig:correlations}
  \end{figure}

The correlation matrix for the timing parameters produced by {\sc tempo} (Figure~\ref{fig:correlations}) indicates only small correlations between orbital and astrometric parameters. The largest correlations are between the three orbital parameters, $x$, $\gamma_{\rm E}$ and $\omega_0$; this explains their lower numerical precision than the other parameters in the timing model [\cite{2019MNRAS.490.3860R}, their equations 23 and 24]. There is also a relatively large correlation between $e$ and $h_3$ (this is expected, see \cite{2010MNRAS.409..199F}, their table 4). There are no additional large ($> 0.9$) correlations between other parameters.

\subsubsection*{Masses from $\chi^2$}

The observed $\dot{P}_\mathrm{b}$ is not caused by gravitational wave damping, and we have only marginal limits for the other PK parameters, so the only parameter that directly constrains $M$ is $\dot{\omega}$. However, the upper limits on $\gamma_{\rm E}$ and $h_3$ provide additional information on the masses. These limits have opposite effects: the upper limit on $\gamma_{\rm E}$ excludes very large values of $m_\mathrm{c}$, while the upper limit on $h_3$ excludes high orbital inclinations and small values of $m_\mathrm{c}$. 

To take all these effects into account, we used the DDGR timing model, a modification of the DD model which directly fits the two masses ($M$ and $m_\mathrm{c}$), assuming that all relativistic effects in the timing of the system are as predicted by GR \cite{DD_1986}. From the pulsar timing observations we only obtain the product $Gm_j$ ($j=c,p$). The masses in units of nominal solar masses $(\Msun)$ are given by the ratio $(Gm_j)/({\cal GM})_\odot^{\rm N}$, where $({\cal GM})_\odot^{\rm N} \equiv 1.3271244 \times 10^{26}\:\mathrm{cm^3\,s^{-2}}$ denotes the nominal solar mass parameter \cite{Prsa_2016}. Because the model returns this ratio, the measurement uncertainty in $G$ cancels.
In cases where absolute masses are needed in physical units we use the value $G = 6.6743 \times 10^{-8}\:\mathrm{cm^3\,g^{-1}\,s^{-2}}$ for conversion \cite{Prsa_2016}. 

As discussed above, the variation of the orbital period ($\dot{P}_\mathrm{b}$) in Table~\ref{tab:ephemeris} is not caused by gravitational wave damping but instead by the system's acceleration in the gravitational field of NGC~1851. This is taken into account in the model by the inclusion of an extra free parameter $\Delta \dot{P}_\mathrm{b}$, which quantifies non-relativistic contributions to $\dot{P}_\mathrm{b}$.

\begin{figure}[H]
    \centering
    \includegraphics[scale=0.15]{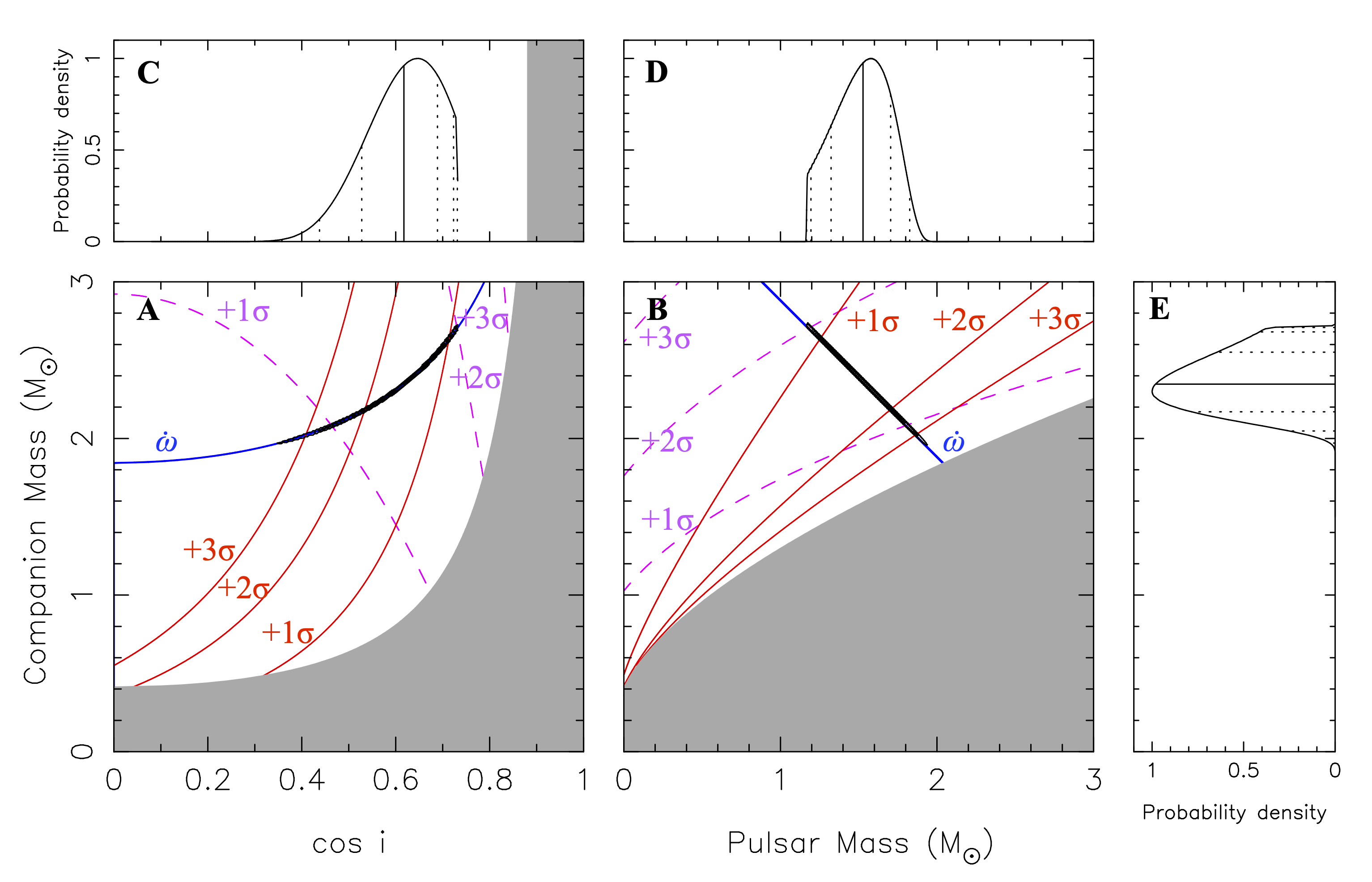}
    \caption{\textbf{Mass and orbital inclination ($i$) constraints for the \E\ system}. In panel \textbf{(A)}, we display the constraints on $\cos i$ and the companion mass $m_\mathrm{c}$; the gray shaded area is excluded by $m_\mathrm{p} > 0$. In panel \textbf{(B)}, we display the same constraints on $m_\mathrm{p}$ and $m_\mathrm{c}$, the gray area is excluded by the mass function and $\sin i \leq 1$. The solid blue lines indicate the nominal and $\pm 1$-$\sigma$ constraints on the total mass of the binary ($M$) obtained from the measurement of the rate of advance of periastron ($\dot{\omega}$). At the scale of the figure, these three lines appear practically superposed. The solid red and dashed magenta lines indicate upper 1, 2 and 3-$\sigma$ limits on $h_3$ and $\gamma_{\rm E}$ respectively. The thick black lines are contours that include 95.4\% of the total probability density. Panels \textbf{(C), (D)} and \textbf{(E)} present the probability density functions (pdfs) for $\cos i$, $m_\mathrm{p}$ and $m_\mathrm{c}$ respectively, these are normalized to the point of maximum probability density. The black lines indicating the median (solid) and the limits of the confidence intervals with equivalent $\pm$ 1, 2 and 3-$\sigma$ confidence levels (dashed). These pdfs are truncated below $m_\mathrm{p} = 1.17 \, \Msun$, the smallest known NS mass.}
    \label{fig:mass-mass}
\end{figure}

We obtain $M \, = \, 3.887 \pm 0.004 \, \Msun$ and  $m_\mathrm{c} \, = \, 2.44 \pm 0.46 \, \Msun$; this implies  $m_\mathrm{p} \, = \, 1.44 \pm 0.46 \, \Msun$. The uncertainties are not Gaussian distributed; the 2-$\sigma$ upper limit for $m_\mathrm{p}$ is above the physical upper limit for $m_\mathrm{p}$ derived from $M$ and the mass function (Figure~\ref{fig:mass-mass}). By forcing \textsc{tempo} to iterate several times, we find that there is no stable numerical convergence on a particular value of $m_\mathrm{c}$, because the relativistic effects used to estimate $m_\mathrm{c}$ are not detected.

To estimate uncertainties and take additional relativistic effects into account in a self-consistent way, we have used a Bayesian technique described in detail elsewhere \cite{2002ApJ...581..509S}, with some adjustments.

For each point in a two-dimensional (2-D) grid of values of $M$ and $\cos i$, we calculate $m_\mathrm{c}$ using equation~\ref{eq:mass_function}, then use $M$ and $m_\mathrm{c}$ as fixed parameters in a DDGR timing model. We  use {\sc tempo} to fit this model to the timing data set (a set of ToAs, $\{ t_j  \}$); this is done by minimizing the weighted sum of the squares of the residuals by varying all parameters in the model apart from $M$ and $m_\mathrm{c}$ (astrometric and spin parameters, Keplerian orbital parameters and the extra $\dot{P}_\mathrm{b}$ caused by the acceleration of the system relative to that of the Solar System projected along the line of sight). For each point in the grid, {\sc tempo} returns a $\chi^2$ value corresponding to the goodness of fit to $\{ t_j  \}$, which we denote $\chi^2(M, \cos i)$.

If $\chi^2_{\rm min}$ is the global minimum of $\chi^2(M, \cos i)$, then each value of
\begin{equation}
\Delta \chi^2 (M, \cos i ) = \chi^2(M, \cos i) - \chi^2_{\rm min}
\end{equation}
has a $\chi^2$ distribution with two degrees of freedom \cite{2002ApJ...581..509S}. We map this to a Bayesian likelihood function $p ( \{ t_j \} | M, \cos i)$:
\begin{equation}
p( \{ t_j \} | M, \cos i) = \frac{1}{2} e^{-\frac{\Delta \chi^2}{2}}.
\label{eq:likelihood}
\end{equation}
The 2-D joint posterior probability density function (pdf) for $M$ and $\cos i$, $p(M, \cos i | \{ t_j \})$, is given by Bayes' theorem:
\begin{equation}
p(M, \cos i | \{ t_j \}) = \frac{p( \{ t_j \} | M, \cos i)}{p( \{ t_j  \} )} p(M, \cos i),
\end{equation}
where $p( \{ t_j  \} )$ is the Bayesian evidence, which is calculated from the integral of $p(M, \cos i | \{ t_j \})$ over areas of the parameter space where $m_\mathrm{p} \equiv M - m_\mathrm{c} > 1.17 \, \rm M_{\odot}$ \cite{Martinez_2015,Suwa_2018}.

The quantity $p(M, \cos i)$ is the Bayesian prior; for these we chose a uniform distribution in $\cos i$ because that corresponds to assuming that there is no favoured prior orientation to the orbital angular momentum. However, there is no need to sample the full $\cos i$ space: for the median value of $M$ the constraint on $m_\mathrm{p}$ implies that $0 < \cos i < 0.732$, thus $90^\circ > i >  42.9^\circ$, so we considered a uniform distribution of $\cos i$ only within this interval. We do not sample the region $-1 < \cos i < 0$ because it cannot be distinguished from the regions between $0$ and $1$ given the available timing precision.
We also assume a uniform prior distribution of $M$, due to our prior ignorance about the total mass of the system. We work with $M$ (not the individual masses) because the measurement of $\dot{\omega}$ is directly related to $M$, assuming GR. A different choice of prior (such as a uniform distribution of $m_\mathrm{c}$) would require sampling much larger regions of parameter space, much of which would have low probability densities, greatly increasing the computational effort without substantially changing the results.

We marginalize 2-D joint posterior pdfs for $M$, $\cos i$, $m_\mathrm{c}$ and $m_\mathrm{p}$. Because $m_\mathrm{c}$ and $m_\mathrm{p}$ are known for each point of the $M$-$\cos i$ grid, we update their marginalized posterior pdfs after calculating $p(M, \cos i | \{ t_j \})$, as we do for $M$ and $\cos i$. At the end of the process, these marginalized posterior pdfs are normalized.

The medians of the marginalized posterior pdfs and equivalent 1-$\sigma$ confidence intervals are presented in the ``Nature of the companion'' section of the main text. The upper and lower limits of the confidence intervals are calculated in such a way that they include 34.13\% of the total probability above and below the medians of each pdf. For a 2-$\sigma$ equivalent confidence level (95.45 \%) around the medians, we obtain the following confidence intervals: $M = 3.887 \:\pm\: 0.009 \: \Msun$, $m_\mathrm{p} = 1.53^{+0.30}_{-0.33} \, \Msun$, $m_\mathrm{c} = 2.35^{+0.33}_{-0.30}\, \Msun$ and $i = 52_{-8}^{+12 \, \circ}$. The mass of $m_\mathrm{c}$ in this estimate is lower than the value estimated in the DDGR model, it is more conservative because the prior (a flat distribution in $\cos i$) makes higher orbital inclinations more likely, resulting in lower values of $m_\mathrm{c}$ (equation~\ref{eq:mass_function}). The choice of sampling region is also conservative: if we included pulsar masses below $1.17 \, \rm M_{\odot}$, then we would be sampling regions with very large values of $m_\mathrm{c}$, which would increase the median for the latter quantity.

\subsubsection*{Search for plasma from the companion}

\begin{figure}[H]
    \centering
    \includegraphics[width=\linewidth]{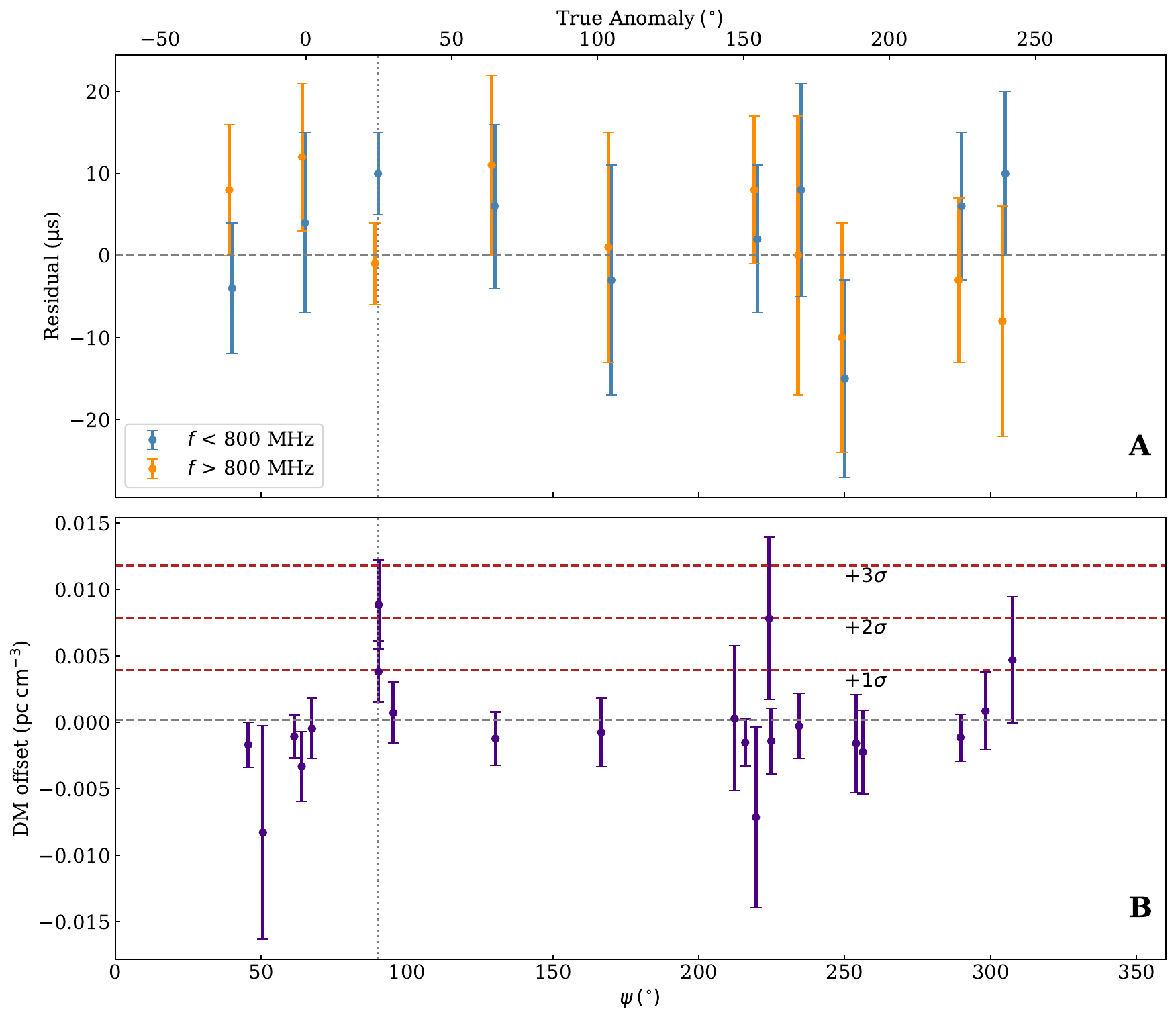}
    \caption{\textbf{Timing residuals and dispersion measure offset as a function of angular orbital position.} \textbf{(A)} Averaged residuals between the ToAs in the MeerKAT UHF band the and timing model of \E\, split into two frequency bands as a function of angular orbital position $\mathrm{\psi}$ with respect to the ascending node of the pulsar. \textbf{(B)} Same as panel A, but for DM offsets (with respect to the pulsar's DM (Table~\ref{tab:ephemeris})). An orbital position of $\mathrm{\psi}=90^\circ$ corresponds to superior conjunction and $f$ denotes frequency.}
    \label{Hint_of_gas_check}
\end{figure}

If the pulsar's binary companion is a main-sequence star, it is likely that plasma expelled from the companion fills a large fraction of the intra-binary medium. For a 2.3~$\Msun$ main-sequence companion and a 1.6~$\Msun$ pulsar, we calculate the Roche-lobe radius [using \cite{Eggleton_1983}, their equation~(2)] at periastron is 3.027~$\Rsun$, using the nominal solar radius adopted by the International Astronomical Union \cite{Prsa_2016}. The corresponding Roche-lobe filling factor is 0.536. The filling of the intra-binary medium due to the plasma from the companion leads to two potentially observable effects: firstly, there could be eclipses of the pulses from the pulsar if the intervening plasma is clumpy. We do not find any evidence for any eclipses. Secondly, any intervening plasma would contain free electrons which the radio signal must pass through; this would cause changes in pulse dispersion at different points in the orbit. We investigate this in two ways: firstly, we average the timing residuals from observations recorded at similar angular orbital positions with respect to the ascending node ($\mathrm{\psi}$), independently for the top and bottom half of the MeerKAT UHF band, to measure if there are any residual frequency-dependent trends in the ToAs. We do not find any evidence for this. Secondly,  we use multi-frequency ToAs for each observation, and independently measure the DM per epoch. Figure \ref{Hint_of_gas_check} shows the averaged timing residuals and DM offset measurements ($\delta\,$DM; after subtracting the nominal DM of the pulsar, 51.93 $\mathrm{pc}$ $\mathrm{cm}^{-3}$) as a function of the orbital position $\mathrm{\psi}$ measured from the longitude of the ascending node. The position in the orbit when the companion is located between the observer and the pulsar along the line of sight is defined as superior conjunction, which is at an angular orbital position of 90 degrees with respect to the ascending node. We do not see any statistically significant ($>3 \sigma$) deviations in the variation of DM along the orbit. 

The second potential indication of intra-binary plasma is variable Faraday rotation, caused by a magnetic field along the line of sight, along the orbit. This would produce an orbital phase dependent change of the rotation measure of the pulsar. Only 17 of the 24 MeerKAT observations include full polarisation information; of these only 3 detect the RM with {$>3\sigma$} significance, due to the low signal-to-noise  and low polarisation fraction \cite{2022A&A...664A..27R} of the pulsar. The RMs of these observations are $13.5 \pm 2.5$, $10.7 \pm 2.0$ and $12.3 \pm 5.7\,  \rm rad \, m^{-2}$. These are consistent with each other, and have an weighted average of $12.2 \pm 2.2\,  \rm rad \, m^{-2}$. We find no evidence for variable RM.

   
\subsubsection*{Search for radio pulsations from the companion}

If the companion star is a NS, it could potentially produce observable pulsed radio emission under the conditions that it is a radio pulsar, its beam sweeps across the Earth, and its average pulsed flux density is above the detection limit of our data. No companion pulsar was detected in the earlier GC pulsar survey \cite{2022A&A...664A..27R}. We performed an additional search specifically targeted on \E\ by de-dispersing each of the 24 MeerKAT observations of NGC 1851 at the DM  of \E\ (51.93 $\mathrm{pc}$ $\mathrm{cm}^{-3}$), using \textsc{presto}'s \texttt{prepdata} routine. We ran the latter in combination with radio frequency interference (RFI) masks, to filter out artificial signals from the data.

For observations that covered an orbital phase interval far from periastron, we performed an acceleration search using the \texttt{accelsearch} routine, with a ``zmax'' parameter of 50. At these orbital phases, we assume the line-of-sight acceleration undergone by the companion (and the pulsar) is constant across the 2 to 4 hours of each observation, hence we expect the acceleration search to recover the Fourier power of any pulsed signal from the companion.
This assumption does not hold for the five observations that were taken around the periastron passage, for which we expect the line-of-sight acceleration to change substantially during the 2 to 3 hours of the observations. The expected change is so rapid that even a search for jerk (rate of change of acceleration) would not be effective.
To recover the Fourier power associated with potential companion pulsations from the five observations taken around periastron, we used the \textsc{pysolator}\cite{pysolator_url} software, which removes the orbital modulation affecting the signal of the pulsar and companion, using the orbital parameters of the latter.
In the case of the \E\ system, we do not know the size of the projected semi-major axis of the companion orbit, $x_\mathrm{c}$, because this depends on the (known) projected semi-major axis of the pulsar orbit, $x_\mathrm{p}$, and the unknown mass ratio $q=m_\mathrm{p}/m_\mathrm{c}$, via the relation $x_\mathrm{c} = q\, x_\mathrm{p}$. Therefore, we demodulated the companion orbit assuming trial $q$ values in the range between 0.519 to 0.784 (which reflects the 1-$\sigma$ uncertainties on $m_\mathrm{p}$ and $m_\mathrm{c}$) with a step size of 0.03. As the trial $q$ value approaches the underlying true mass ratio of \E, the effect of the orbital motion on the companion's signal is minimised, allowing searches over a range of constant accelerations to become effective. For each periastron passage observation, we performed an acceleration search with a ``zmax'' parameter of 20, on all the demodulated time series. All candidate pulsar signals produced from the 24 observations were folded modulo the detected spin frequency with \textsc{presto}'s \texttt{prepfold} and the resulting diagnostic plots were inspected visually. We recovered several of the 15 pulsars known in NGC 1851 with DMs close to that of \E \cite{2022A&A...664A..27R}. The known pulsars in the cluster are all ruled out as the companion, as they are not spatially coincident with and do not share the same orbital parameters as \E. We found no pulsar-like signal that could be ascribed to the companion.

\subsection*{Near-ultraviolet and optical observations and data analyses}
\label{sec:hst_obs}

\subsubsection*{Hubble Space Telescope observations}

We utilised archival observations of NGC 1851 with Wide Field Camera 3 (WFC3) on the Hubble Space Telescope (HST) taken in 2010 as part of the observing programme with proposal ID 12311 (Principal Investigator: Piotto) and in 2014 through proposal ID 13297 (Principal Investigator: Piotto). We downloaded from the MAST archive observations in the F275W and F336W filters, which are sensitive to hot stars such as blue stragglers and WDs. The data-set is composed of 14 images acquired with the F275W filter with exposure time of 1280 s, and 4 images acquired with the F336W filter with exposure time of 453 s.  

\subsubsection*{Photometric analysis}

The photometric analysis was performed using \textsc{DAOPHOT} \cite{1987PASP...99..191S} following methods described elsewhere \cite{2019ApJ...875...25C,2020ApJ...905...63C}. First, we selected a few hundred bright stars to model the point spread function (PSF) of each image. These PSF models were applied to all the sources detected at more than $5\sigma$ above the local background level. Then, we produced a reference list of stars including all the sources detected in at least half the F275W images. At the corresponding positions of these stars, PSF fitting was forced in all the other frames using 
the \texttt{allframe} routine in \textsc{DAOPHOT} \cite{1994PASP..106..250S}. For each star we homogenized and averaged the magnitudes estimated in different images using 
the \texttt{daomaster} routine in \textsc{DAOPHOT}, to obtain the instrumental stellar magnitudes and uncertainties. 

The observed magnitudes were calibrated to the Vega photometric system by cross-correlation with a previous catalogue derived from the same observations of NGC~1851 \cite{2015AJ....149...91P}. 

The instrumental positions were corrected for geometric distortion following previously described methods \cite{2011PASP..123..622B} and converted to the ICRS using the stars in common with the Gaia Data Release 3 (DR3) catalog \cite{GaiaDR3}. 
We used the cross-correlation software \textsc{CataXcorr}\cite{cataxcorr_url} adopting a six parameter linear transformation to convert the instrumental pixel positions to the ($\alpha$, $\delta$) absolute coordinate system using Gaia DR3 stars as reference. The residuals of this transformation showed a combined root mean square of 14 mas (9 mas in each coordinate), which we adopt as the $1\sigma$ astrometric uncertainty.

\subsubsection*{Stars near the position of \E}

Two stars are observed within 100 mas of the position of \E.
As discussed in the main text, we identify one as a blue straggler ($m_{F275W}=18.575\pm0.009$, $m_{F336W}=18.15\pm0.02$) and the other as a red giant ($m_{F275W}=19.047\pm0.007$, $m_{F336W}=17.60\pm0.02$); both of which have estimated masses below the lower limit on the mass for the companion of \E\, ($1.4 \, \Msun$) obtained without assuming that $\dot{\omega}$ is relativistic. Assuming that the blue straggler star (BSS) is a H-core burning star, we estimate its mass through a theoretical mass-luminosity relation \cite{2008ApJS..178...89D} assuming the cluster metallicity, distance and reddening \cite{harris2010} (2010 edition). We estimate that the blue straggler's magnitude is compatible with a $\sim 1.2 \, \Msun$ star, too low to be the companion of \E. 
Direct estimation of the mass of the BSS is not possible with the available observations. Main sequence mass-luminosity relationships can be used to estimate BSS masses \cite{2014ApJ...783...34F,2019ApJ...879...56R}, though with a higher uncertainty on the BSS mass of $0.5\,\Msun$.

We found no evidence of photometric variability in either of the two stars near the position of \E. If in a binary, variability could arise from heating or eclipsing of the companion star. Although the observations sample only $\sim25\%$ of the binary orbital period, the lack of variability is consistent with our interpretation that neither of the two stars is associated with the pulsar.

\section*{Supplementary Text}

\subsection*{Formation of massive compact binaries in NGC 1851}

Although NGC~1851 is not a core-collapsed GC, it does have a high central density relative to most other GCs [\cite{BaumgardtHilker18}, their table 2]. This and its small core radius of $0.09'$, which was determined using the stellar density profile \cite{Miocchi_2013}, imply that it has a high rate of stellar interactions per binary ($\gamma$) [\cite{2014A&A...561A..11V}, their figure 1]. The high theoretical $\gamma$-value of NGC~1851 is supported by observations of three highly eccentric MSPs in binaries in this GC with degenerate companions that are much more massive than the companions associated with similar systems in the Galactic disk: PSR~J0514$-$4002A \cite{2004ApJ...606L..53F,2007ApJ...662.1177F,2019MNRAS.490.3860R}, PSR~J0514$-$4002D and \E \cite{2022A&A...664A..27R}.

The encounter rate per binary affects the size and composition of the binary population: a GC with a large overall stellar interaction rate (like for instance, 47~Tuc) will form large numbers of LMXBs and subsequently radio MSPs, but if the $\gamma$-value is low, these binaries will likely evolve without disturbance. Most binary MSP systems found in low-$\gamma$ GCs have low-mass companions (WDs) in near-circular orbits \cite{2017MNRAS.471..857F, Wang_2020} --- i.e. they resemble the MSP population in the Galaxy \cite{2023pbse.book.....T}. Such GCs are expected to form tight DNS systems at rates that are much smaller than the observed rate of DNS mergers detected by gravitational waves, which is supported by simulations \cite{yfk+20}. On the other hand, in GCs with a high $\gamma$-value, several subsequent exchange encounters involve the same MSP, after the exchange that formed its parent LMXB, which could in principle form many DNS systems. This is the case for NGC~1851 \cite{yfk+20}. If the merger products of such DNSs stay in the cluster (i.e. if the merger kick is not larger than the escape velocity of the cluster), they will be available for subsequent exchange interactions. 

We compare the expected merger kick imparted on a DNS merger product to the escape velocity of NGC~1851. Following previous calculations [\cite{2021MNRAS.502.2049L}, their equation~10], we find  a $1.2+1.4\;M_\odot$ DNS merger has an estimated merger kick of $w_{mk}=43.8\;{\rm km\,s}^{-1}$, which is close to the central escape velocity of NGC~1851 of $42.9\;{\rm km\,s}^{-1}$ \cite{BaumgardtHilker18}. However, the closer the unknown mass ratio, $q$, of the two merging NSs is to unity, the smaller is the imparted merger kick ($w_{mk}\rightarrow 0$ as $q\rightarrow 1$), and the higher the probability is for the merger remnant to remain within the GC. Hence, we conclude that it is no problem to keep a DNS merger remnant within NGC~1851 from a kinematic point of view.

Dense GCs experience mass segregation, in which more massive stellar populations concentrate closer to the centre, especially in the positions of pulsars and X-ray sources \cite{2005ApJ...625..796H}. Therefore, the most massive compact objects (NSs, black holes and merger remnants) should, when in equilibrium with the remaining stars in the cluster (when all types of objects have the same average kinetic energy), occupy a small volume near the centre of the GC. Therefore, in the centres of GCs, massive stellar remnants are expected to constitute most of the mass \cite{BaumgardtHilker18}. The measured position of \E\, (Table \ref{tab:ephemeris}) is offset from the optical centre of the cluster \cite{Gaia_Collaboration_2018} by $0.064'$, which, given estimates of the cluster core radius of $0.09'$ and $0.044'$ \cite{Miocchi_2013, BaumgardtHilker18}, places this pulsar either in or close to the core. This is also true of the other massive binary pulsar known in this GC, PSR~J0514$-$4002A. The abundance and close proximity of massive degenerate objects in this region implies a higher probability of formation of systems like  \E. 

\subsection*{Contributions to the observed post-Keplerian parameters}
\subsubsection*{Contributions to $\dot\omega_{\mathrm{obs}}$} \label{subsection} 

The observed rate of advance of periastron for a binary pulsar can be produced by the combination of classical contributions and relativistic effects. The observed change in the longitude of periastron ($\dot\omega_{\mathrm{obs}}$) is \cite{Handbook_Lorimer_Kramer}
\begin{equation}
    \dot\omega_{\mathrm{obs}} = \dot\omega_{M} + \dot\omega_{\mathrm{PM}}+ \dot\omega_{\mathrm{3rd-body}} + \dot\omega_{\mathrm{SO}} \label{omdot_1},
\end{equation}
where $\dot\omega_{M}$ is the leading-order relativistic periastron advance, $\dot\omega_{\mathrm{PM}}$ and $\dot\omega_{\mathrm{3rd-body}}$ are the impact of the proper motion of the system and the possible presence of a nearby third body respectively, and $\dot\omega_{\mathrm{SO}}$ includes the contributions from relativistic and classical spin-orbit coupling. Below, we consider each of these effects in turn.
\par
The relativistic periastron advance can be expressed as a function of the total mass of the system and the measured Keplerian parameters, to the lowest post-Newtonian order and for negligible spin contributions, by re-arranging equation~\ref{eq:Mtot}:
\begin{equation}
    \dot\omega_{M} = \frac{3}{c^2(1-e^2)}\left({\frac{P_\mathrm{b}}{2\pi}}\right)^{-5/3}\left(GM\right)^{2/3}.
\label{omdot_rel}
\end{equation}
\par
A proper motion of the binary would change the longitude of periastron due to the variations in the orientation of the orbital plane with respect to our line of sight. This effect can be quantified as \cite{Kopeikin_1996, Freire_2011a}
\begin{equation}
    \dot{\omega}_\mathrm{PM} = 2.78 \times 10^{-7} \, \frac{\mu_{\rm{T}}}{\sin{i}} \, \cos\left[\Theta_{\mu} - \Omega\right ]\:\mathrm{deg}\,\mathrm{yr}^{-1},   \label{omdot_PM}
\end{equation}
where $\mu_{\rm{T}}$ is the magnitude of the total proper motion in $\mathrm{mas\:yr^{-1}}$, $\Theta_{\mu}$ its position angle, and $\Omega$ is the longitude of the ascending node of the pulsar's orbit. The maximal contribution from $\dot\omega_{\mathrm{PM}}$ is
\begin{equation}
     \dot{\omega}_\mathrm{PM,max} = 2.78 \times 10^{-7} \frac{\mu_{\rm{T}}}{\sin{i}} \:\mathrm{deg}\,\mathrm{yr}^{-1}.
     \label{omdot_PM_max}
\end{equation}
As discussed above, we cannot measure the proper motion of the pulsar, so have assumed the cluster's proper motion value for our timing solution. However, to estimate the maximum contribution from $\dot{\omega}_\mathrm{PM}$, we use a sum of the proper motion of the cluster ($\mu_\mathrm{GC}$) and the proper motion corresponding to the cluster's central escape velocity ($\mu_\mathrm{esc}$). The value of $\sin{i}$ used in this calculation corresponds to the angle of inclination predicted by the timing model. 

The dynamics of the cluster \cite{2022ApJ...934..150L} indicate $\mu_{\mathrm{esc}} = 0.78 \, \rm mas \, yr^{-1}$; adding this to the proper motion of the GC ($\mu_{\mathrm{GC}} = 2.22 \, \rm mas \, yr^{-1}$), we derive an upper limit to the proper motion of \E\ as ${\mu_{\rm{T}}} \leq 3.00 \, \rm mas \, yr^{-1}$. Using this value in equation~\ref{omdot_PM_max}, the maximum contribution to the change in the longitude of periastron from the proper motion of the system can be estimated:
\begin{equation}
    \dot{\omega}_\mathrm{PM,max} = 1.104 \times 10^{-6} \:\mathrm{deg}\,\mathrm{yr}^{-1}. \label{omd_PM_maxval}
\end{equation}
The corresponding contribution to the observed rate of advance of periastron,
\begin{equation}
    \dot{\omega}_\mathrm{PM,max} = 3.184 \times 10^{-5} \, \dot\omega_{\mathrm{obs}}, \label{omdotPM/omdotObs}
\end{equation}
is negligibly small. 
\par
\begin{figure}[H] 
  \centering \includegraphics[width = 0.95\textwidth]{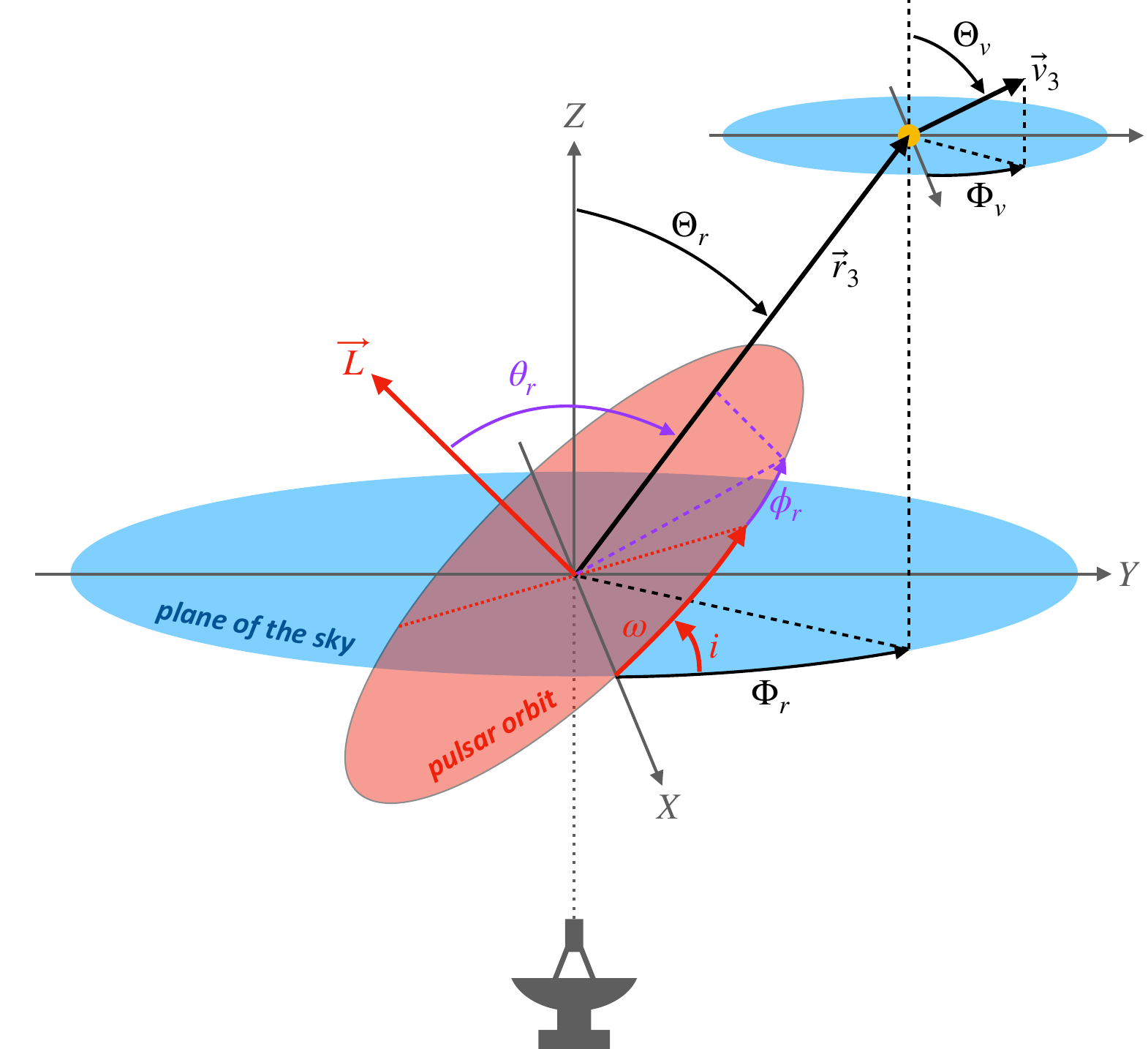}
  \caption{\textbf{Geometry for third-body calculations}. $(X,Y,Z)$ denotes the right-handed Cartesian coordinate system, where the X-axis is aligned with the ascending node of the pulsar orbit and the Z-axis is normal to the plane of the sky. $i$ is the angle of inclination between the pulsar orbit and the plane of the sky, and $\omega$ is the longitude of periaston. The putative third body (orange) is at distance $\vec{r}_3$ and moves with velocity $\vec{v}_3$. We assume uniform probability distributions for $\Phi_r$, $\cos\Theta_r$, $\Phi_v$, and $\cos\Theta_v$. The angles $\phi_r$ and $\theta_r$ are defined with respect to the orbit ($\vec{L}$: orbital angular momentum of the pulsar binary) [\cite{Joshi_Rasio_1997}, their $\phi_2$ and $\theta_2$ respectively].}
  \label{fig:MCgeom}
\end{figure}

 
The presence of an additional third body near the system (either bound to it, or more likely in a close flyby) could also contribute to the observed change in the longitude of periastron. However, we expect a third body that contributes to the $\dot\omega_{\mathrm{obs}}$ to cause additional secular changes in other binary parameters (especially orbital inclination and eccentricity), and produce spin-frequency derivatives ($\dot\nu$, $\ddot\nu$, $\dots$) that are orders of magnitude larger than we observe. We assume the geometry illustrated in Figure~\ref{fig:MCgeom}. Neglecting angular dependencies and corrections due to eccentricity, $|\dot\omega_\mathrm{3rd-body}| \sim (3\pi/P_\mathrm{b})(m_3/M)(a_\mathrm{pc}/r_3)^3$ [\cite{Joshi_Rasio_1997}, their equation~(33)], where $m_3$ is the mass of the third body and $a_\mathrm{pc}$ denotes the semi-major axis of the relative motion of the (inner) binary system (about $0.12$ astronomical units (au), derived using Kepler’s third law). The minimum distance $r_3$ for an external mass $m_3$ to produce a $\dot\omega_\mathrm{3rd-body}$ that has a non-negligible contribution to the $\dot\omega_{\mathrm{obs}}$ (more than two times its measurement uncertainty), and the gravitational acceleration consequently caused by this third body, can be determined using this equation. Using the values in Table~\ref{tab:ephemeris} and equation~(3) in \cite{Joshi_Rasio_1997} we find
\begin{equation} \label{eq:nudotext}
    \left|\frac{\dot\nu_\mathrm{3rd-body}}{\dot\nu}\right| \gtrsim 180 
    \left(\frac{m_3}{\mathrm{M}_\odot}\right)^{1/3} |\cos{\Theta_r}| \,.
\end{equation}
Consequently, even a small mass $m_3$ has to be close to the plane of the sky ($\Theta_r = \pi/2$) to produce the observed $\dot\nu$. $m_3$ cannot be very small, otherwise the third body would have to be very close to the (inner) system, which would produce higher order derivatives in $\nu$ \cite{Joshi_Rasio_1997}. We use an equivalent analysis to obtain an expression for $\ddot\nu$ [\cite{Joshi_Rasio_1997}, their equation~(3)]. We assume a (bound) circular orbit for the third body and leading order in $m_3/M$, leading to
\begin{equation} \label{eq:nu2dotext}
    \left|\frac{\ddot\nu_\mathrm{3rd-body}}{\ddot\nu}\right| \gtrsim 14000
    \left(\frac{m_3}{\mathrm{M}_\odot}\right)^{-1/6} |\cos{\Theta_v}| \,.
\end{equation}
Consequently, the motion of the third body again has to be (very) closely aligned with the plane of the sky ($\Theta_v \simeq \pi/2$) to explain the measurement. In our example, the combination of equations \ref{eq:nudotext} and \ref{eq:nu2dotext} requires a nearly perfect alignment of the orbital plane of the putative third body with the plane of the sky. This is a highly unlikely orientation (probability of $\lesssim$ $10^{-5}$, before accounting for the probability of having a third body in sufficient proximity ($\lesssim (57\,{\rm au})(m_3/\Msun)^{1/3}$)). We expect such a third body would produce related changes in the eccentricity and projected semi-major axis of the pulsar orbit [\cite{Joshi_Rasio_1997}, their equations 34 and 35]. 

We may make this result more rigorous by removing the assumption of circular motion for the third body (outer orbit) and accounting for the high eccentricity of the inner orbit. We have obtained the corresponding expressions for the tidally induced $\dot{\omega}$, $\dot{e}$ and $\dot{x}$ by (semi-analytically) integrating the equations of osculating orbital elements, using as a perturbing force the differential acceleration in the binary system caused by the external mass $m_3$ [see \cite{PoissonWill_2014}, their equations 3.69 and 3.77] (note, the fully analytical equations for the tidally induced $\dot\omega$, $\dot{x}$ and $\dot{e}$ given in \cite{Joshi_Rasio_1997} assume a small eccentricity for the inner orbit which is not the case in the \E\, system, where we measure $e \approx 0.71$). These calculations use the geometry of Figure~\ref{fig:MCgeom}, and allow the calculation of the tidally induced $\dot\omega$, $\dot{x}$ and $\dot{e}$ for any given distance $r_3 = |\vec{r}_3|$ and given directional angles $\Phi_r$ and $\Theta_r$. Like for the circular outer orbit above, we need the first and second frequency derivative, while this time however we allow the velocity of the third body to have any magnitude and direction (see Figure~\ref{fig:MCgeom}). Like before, we used the standard equations for the externally induced time derivatives of the pulse frequency, $\dot\nu_\mathrm{3rd-body}$ and $\ddot\nu_\mathrm{3rd-body}$ [\cite{Joshi_Rasio_1997}, their equation~(3)]. 

\begin{figure}[H] 
  \centering \includegraphics[width = 0.95\textwidth]{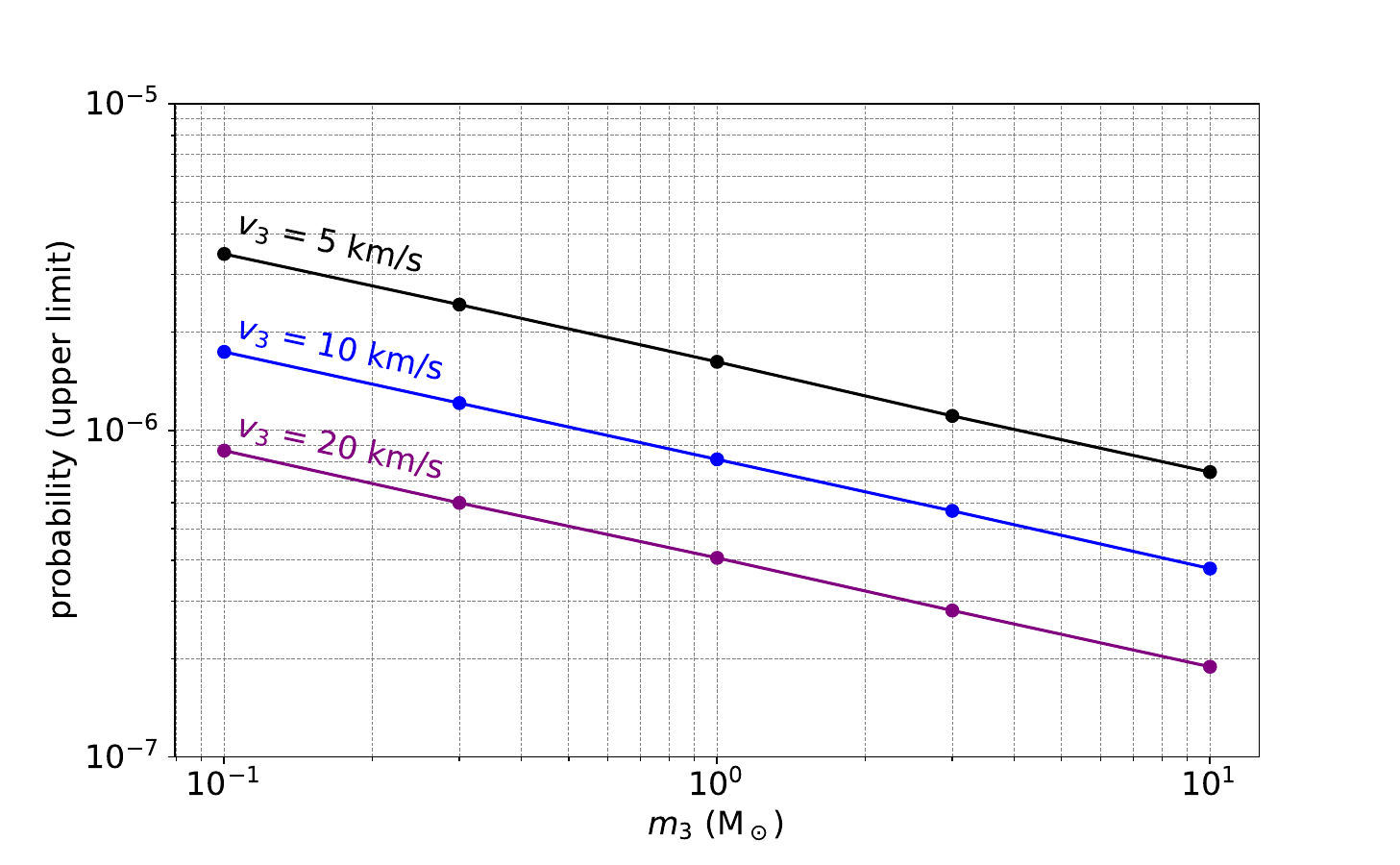}
  \caption{\textbf{Upper limits on the probability of non-negligible tidal contributions to $\dot\omega_\mathrm{obs}$}. The figure shows the upper limit on the probability that $\dot\omega_{\mathrm{obs}}$ has a tidal contribution (caused by a nearby third body) that exceeds twice the measurement uncertainty. Results are shown for different masses ($m_3$) and velocities ($v_3$) of the third body (see text). Masses significantly below 0.1\,$\mathrm{M}_\odot$ are excluded, since they have to be very close to the binary system in order to meet the $\dot\omega_\mathrm{obs}$ condition above. As a consequence, they would lead to higher order time derivatives in the spin frequency ($\dddot{\nu}$,$ \ddddot{\nu}$, \ldots), which are not observed.}
  \label{fig:MCresults}
\end{figure}

In order to estimate an upper limit on the probability for the presence of a third body, we Monte-Carlo (MC) sample the four dimensional parameter space of $\Phi_r$, $\Phi_v$,  $(\cos\Theta_r)$, and $(\cos\Theta_v)$, which all have a uniform probability distribution over their corresponding range ($[0,2\pi)$ for the first two and $[-1,+1]$ for the last two).
When sampling the ($\Phi_r$, $\cos\Theta_r$, $\Phi_v$, $\cos\Theta_v$) parameter space, we keep the distance $r_3$ as a free parameter, to maximise the tidal contribution to $\dot\omega_{\mathrm{obs}}$ (i.e.\ find $\max|\dot\omega_\mathrm{3rd-body}|$) within the constraints given by the observed frequency derivatives (we use $|\dot\nu_\mathrm{3rd-body}| < 3 |\dot\nu_\mathrm{obs}|$ and $|\ddot\nu_\mathrm{3rd-body}| < 3 |\ddot\nu_\mathrm{obs}|$) and orbital parameters (within the 3-$\sigma$ range of $\dot{x}$ and $\dot{e}$). We therefore regard our results as conservative upper limits because they are obtained without considering the low probability of having a third body in the proximity to \E. 
To obtain a high accuracy in our MC sampling of the ($\Phi_r$, $\cos\Theta_r$, $\Phi_v$, $\cos\Theta_v$) parameter space, we conducted $10^{10}$ MC realisations for each individual combination of mass $m_3$ and velocity $v_3$, always counting those configurations that have a $\max|\dot\omega_\mathrm{3rd-body}|$ that is larger than two times the measurement uncertainty of $\dot\omega_{\mathrm{obs}}$, while at the same time pass the spin-frequency ($\dot\nu$, $\ddot\nu$) and orbital ($\dot{x}$, $\dot{e}$) constraints. Our procedure corresponds to calculating a (normalized) volume in the four-dimensional parameter space of ($\Phi_r$, $\cos\Theta_r$, $\Phi_v$, $\cos\Theta_v$), defined by the given boundary conditions. This integral is solved numerically using Monte Carlo integration, a standard technique for higher-dimensional integrals. 
The results of our MC runs are shown in Figure~\ref{fig:MCresults}. Based on these results, we conclude that the probability for a non-negligible (more than two times the measurement uncertainty) tidal contribution to $\dot\omega_{\mathrm{obs}}$ is $\sim10^{-6}$, so only highly fine-tuned (directional) configurations could produce such an effect. The maximum $\dot\omega_\mathrm{3rd-body}$ from our MC runs that does not violate our observed values for $\dot{x}$ ($(4\,\pm\,2) \times 10^{-13}$) and $\dot{e}$ ($(-0.06 \pm 1.11) \times 10^{-14}\,\rm{s}^{-1}$) is $\sim 1\%$ of $\dot\omega_\mathrm{obs}$. This is too low to influence the conclusions of this paper.

We note that $10\,\rm{km}\,\rm{s}^{-1}$ as the relative velocity between the third body and the inner binary corresponds to the escape velocity (lower limit for flyby) at about $100\, {\rm au}$.

\par
A secular change in the longitude of periastron could arise from the contribution of the spins of the pulsar and its binary companion due to Lense-Thirring precession ($\dot\omega_{\mathrm{LT}}$). A change could also be caused by the spin-induced quadrupole moment of the rotating companion ($\dot\omega_{\mathrm{Q}}$). The $\dot\omega_{\mathrm{SO}}$ term from equation~\ref{omdot_1} can be decomposed as 
\begin{equation}
    \dot\omega_{\mathrm{SO}} = \dot\omega_{\mathrm{LT}} + \dot\omega_{\mathrm{Q}}.   \label{omega_dot_SO}
\end{equation}
We discuss them separately and calculate the maximum contributions for different possible binary companions of \E.
\par
The dimensionless measure of the spin angular momentum ($S_\mathrm{c}$) of a body of mass $m_\mathrm{c}$ is \cite{Chandrasekhar_1983, PoissonWill_2014}
\begin{equation}
    \chi_\mathrm{c} = {\frac{c}{G}}{\frac{S_\mathrm{c}}{m_\mathrm{c}^{2}}}. \label{chi_c}
\end{equation}
For a NS with a mass larger than $\sim 1\, \Msun$, the maximum value of the spin parameter is ${\chi_{\mathrm{max}}\sim0.7}$ \cite{Lo_2011}. This is valid for most proposed equation of states (EoSs) and is independent of the mass of the NS. A rotating Kerr BH in general relativity can have spin parameter values going up to 1, mathematically \cite{Chandrasekhar_1983} (astrophysical BHs are generally expected to have spin values below the Thorne limit, 0.998 \cite{Thorne_1974}; however, the maximum possible spin could also reach values closer to 1 depending on the details of the accretion process \cite{Skadowski_Abramowicz_2011}). For a group of objects including any NS or BH, $\chi_\mathrm{c}\,<\,1$.
\newline
According to \cite{Greif_2020}, for a NS mass of $1.4\,\Msun$, the typical moment of inertia of is ${\sim1.5\times10^{45}\:\mathrm{g}\,\mathrm{cm^2}}$; using this, the spin angular momentum of \E\, is then ${\approx\,1.7\times10^{48}\:\mathrm{g}\,\mathrm{cm}^2\,\mathrm{s}^{-1}}$ and the spin parameter is $\chi \approx 0.1$. This is about an order of magnitude smaller than expected for an extremely fast rotating NS or BH companion. We therefore neglect it hereafter.  
\newline
The contribution to $\dot\omega_\mathrm{obs}$ from the Lense-Thirring precession of the compact companion is \cite{Damour_Schafer_1988} 
\begin{equation}
  \dot{\omega}_\mathrm{LT} 
  = \frac{n_\mathrm{b}B_\mathrm{c}}{\sin^{2}{i}}\:
    \left[{(1-3\,{\sin^{2}{i}})(\hat{\mathbf{k}}\cdot\hat{\mathbf{s}}_\mathrm{c}) - \cos{i}(\hat{\mathbf{K}}_0\cdot\hat{\mathbf{s}}_\mathrm{c})} \right] 
  = {n_\mathrm{b}}{B_\mathrm{c}}{f_\mathrm{c}},
\label{omega_dot_LT}
\end{equation}
where $n_\mathrm{b}$ is the orbital frequency, the unit vector $\hat{\mathbf{k}}$ is normal to the orbital plane defined by the orbital angular momentum, $\hat{\mathbf{s}}_\mathrm{c}$ is the unit vector along the spin of the companion, and $\hat{\mathbf{K}}_0$ is the unit vector along the line of sight. The explicit expressions for $B_\mathrm{c}$ and $f_\mathrm{c}$ are given below. Using equation~\ref{chi_c}, $B_\mathrm{c}$ is [\cite{Damour_Schafer_1988}, their equation 5.17a] 
\begin{equation}
B_\mathrm{c} 
= {\frac{4 + 3{m_\mathrm{p}}/{m_\mathrm{c}}}{2{c^2}M(1-e^2)^{3/2}}} \: {n_\mathrm{b}}{S_\mathrm{c}} 
= {\frac{Gm_\mathrm{c}({3M + m_\mathrm{c}})}{2{c^3}M(1-e^2)^{3/2}}} \: {n_\mathrm{b}}{\chi_\mathrm{c}}.
\label{B_c}
\end{equation}

\begin{figure}[H] 
  \centering \includegraphics[width = 0.95\textwidth]{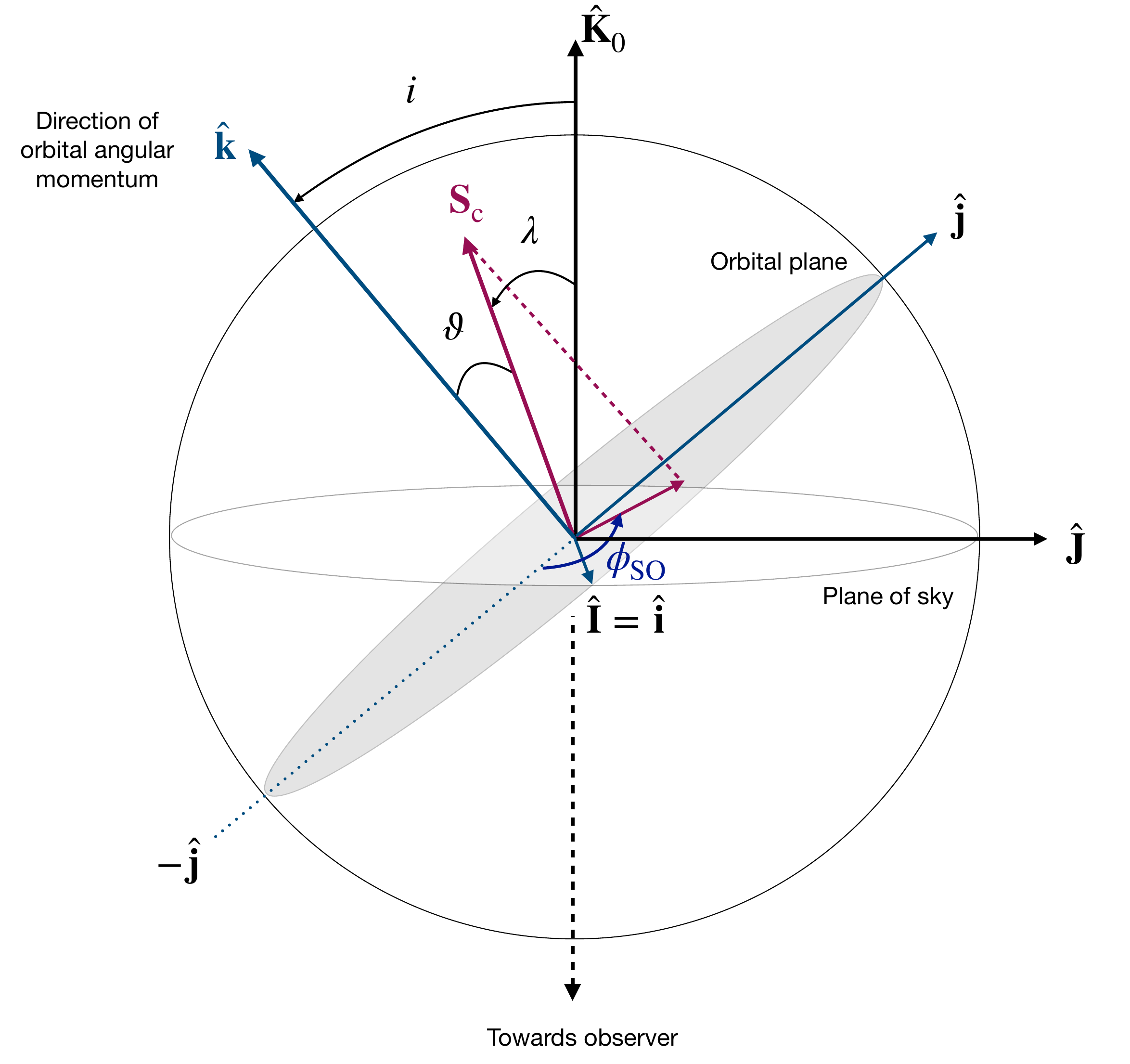}
  \caption{\textbf{Definition of the orbital orientation and geometric parameters of the system.} The conventions of \cite{Damour_Taylor_92} are followed. The plane of the sky is denoted by the vectors $\hat{\mathbf{I}}$ and $\hat{\mathbf{J}}$, and the vectors $\hat{\mathbf{I}} \equiv \hat{\mathbf{i}}$ and $\hat{\mathbf{j}}$ are a part of the orbital plane. The direction of the orbital angular momentum $\hat{\mathbf{k}}$ is perpendicular to the orbital plane and is inclined at an angle $i$ with respect to the line-of-sight vector $\hat{\mathbf{K}}_0$. The spin angular momentum of the binary companion is defined by $\mathbf{S}_c$, and its direction with respect to the line-of-sight is given by the angle $\lambda$. The misalignment angle between the spin and the orbital angular momentum is denoted by $\vartheta$, where $0 \leq \vartheta \leq \pi$. $\mathbf{S}_c$ forms a precession cone around $\hat{\mathbf{k}}$, and the angle swept by the projection of the former in the orbital plane measured from $-\hat{\mathbf{j}}$ is given by $\phi_{\mathrm{SO}}$.} 
  \label{fig:pulsar_figure}
\end{figure}

To estimate the maximum $\dot\omega_{\mathrm{LT}}$ contribution from a NS or BH companion, we calculated the values for $B_\mathrm{c}$ over a range of pulsar and companion masses using $\chi_\mathrm{c}=\chi_{\mathrm{max}}=1$. The factor $f_\mathrm{c}$ can be expressed as
\begin{equation}
f_\mathrm{c}  
= \frac{(1-3\,{\sin^{2}{i}})(\hat{\mathbf{k}}\cdot\hat{\mathbf{s}}_\mathrm{c}) - \cos{i}(\hat{\mathbf{K}}_0\cdot\hat{\mathbf{s}}_\mathrm{c})}{\sin^{2}{i}} 
= -2\cos{\vartheta} + {\sin{\vartheta}}\,{\cot{i}}\,{\cos{\mathrm{\phi}_{\mathrm{SO}}}},    
\label{f_c}
\end{equation}
where $\vartheta$ $(0 \leq \vartheta \leq \pi)$ is the angle between the spin of the companion (${\mathbf{S}}_\mathrm{c} = S_\mathrm{c} \hat{\mathbf{s}}_\mathrm{c}$) and the orbital angular momentum and $\mathrm{\phi}_{\mathrm{SO}} (0 \leq \mathrm{\phi}_{\mathrm{SO}}< 2\pi)$ is the angle swept by the projection of $S_\mathrm{c}$ during its precession around $\hat{\mathbf{k}}$ measured in the orbital plane as shown in Figure~\ref{fig:pulsar_figure} \cite{Damour_Taylor_92}. 

\begin{figure}[H] 
  \centering \includegraphics[width = \textwidth]{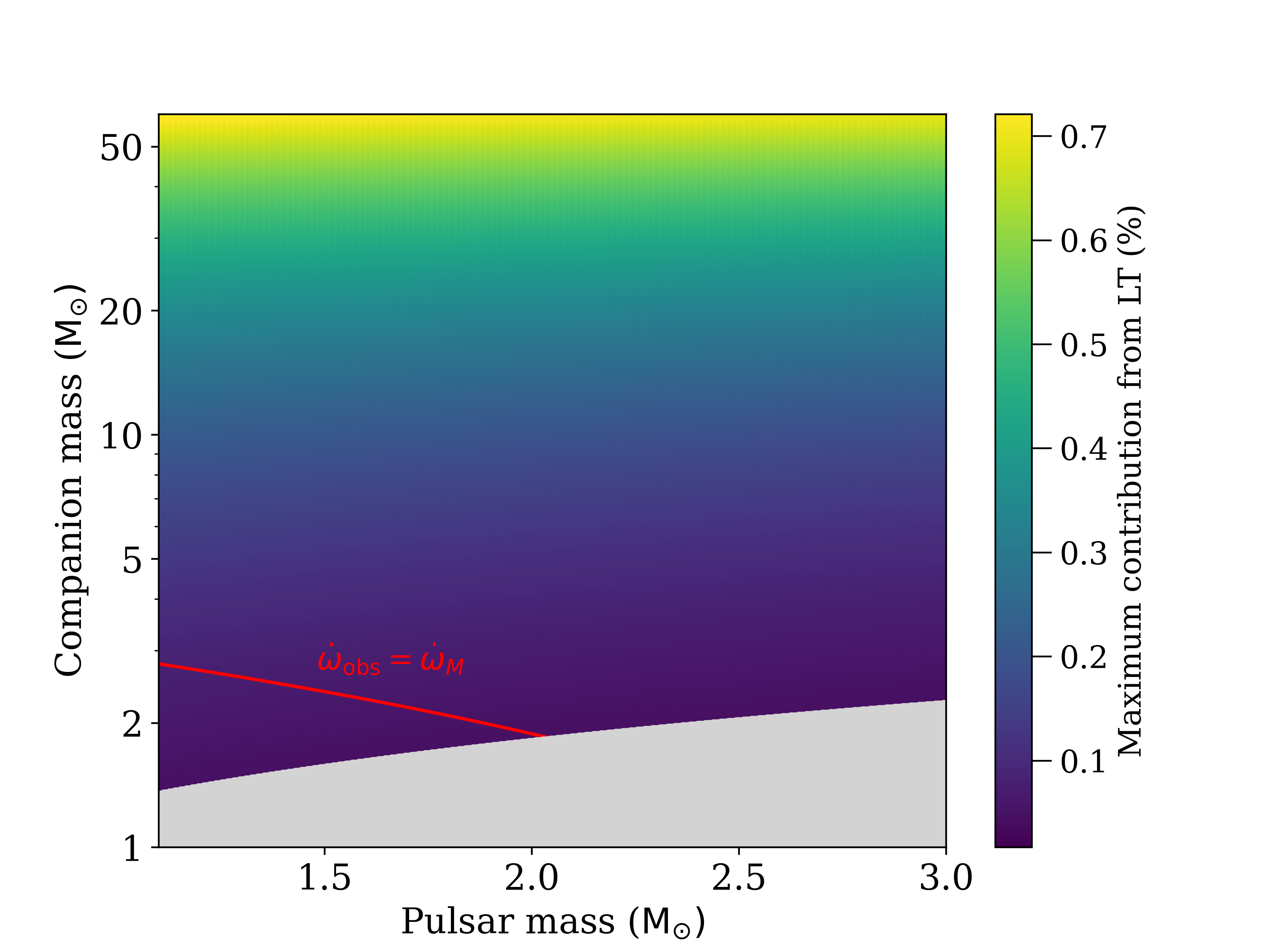}
  \caption{\textbf{Maximum contribution to the rate of advance of periastron from Lense-Thirring precession}. Companion masses are on a logarithmic scale; the upper limit of 60 $\Msun$ is set from radio timing analysis (see text). The pulsar masses, on a linear scale, span 1.1 $\Msun$ to 3.0 $\Msun$ (see text). The grey shaded region is excluded by the constraint $\sin{i} \le 1$ and the boundary of this region corresponds to $i = 90^\circ$. The red curve shows the nominal measurement of $\dot\omega$ for \E, which equals the relativistic rate of advance of periastron given by equation~\ref{omdot_rel}.} 
  \label{fig:omega_dot_LT_variation}
\end{figure}

For our calculations, we used a uniformly distributed array of masses given by $m_\mathrm{p} \in [1.1\,\Msun,$ $4.0\,\Msun]$ and $m_\mathrm{c} \in [1.4\,\Msun, 60.0\,\Msun]$, where the upper limit for the mass of the companion is set as the maximum mass allowed by an offset in the reduced-$\chi^{2}$ value within a 3-$\sigma$ deviation from the solution. A companion with a mass exceeding $60\, \Msun$ would produce a large Shapiro delay, independent of the inclination, inconsistent with our measured limit for the system [\cite{BT_76}, their equation~2.20]. For each pair of pulsar and companion mass $(m_\mathrm{p}, m_\mathrm{c})$ within this range, we evaluate the values of $B_\mathrm{c}$ and $f_\mathrm{c}$ to obtain the maximum $\dot{\omega}_\mathrm{LT}$. The largest absolute value of $f_\mathrm{c}$, for any such given pair, can be expressed as ${\lvert{f_\mathrm{c}}\rvert}^{\mathrm{max}} = \sqrt{4 + \cot^{2}{i}}$. The ratio ($\rho_{\mathrm{LT,1}}$) of the maximum value of $\dot{\omega}_\mathrm{LT}$ with respect to $\dot{\omega}_{M}$ for every pair of mass, 
\begin{equation}
   \rho_{\mathrm{LT,1}} \equiv  \frac{|{\dot\omega}_{\mathrm{LT}}|^\mathrm{max}(m_\mathrm{p},m_\mathrm{c},\chi_\mathrm{c}=1)}{{\dot\omega}_M(m_\mathrm{p},m_\mathrm{c})},
\end{equation}
is shown as a function of mass in Figure~\ref{fig:omega_dot_LT_variation}. The maximum value of this ratio occurs for the largest companion mass and the smallest pulsar mass, reaching
\begin{equation}
   \max(\rho_{\mathrm{LT,1}}) = 0.0072.
\end{equation}
Consequently, for the whole range of masses, i.e.\ $m_\mathrm{p} \in [1.1\,\Msun, 3.0\,\Msun]$ and $m_\mathrm{c} \in [1.4\,\Msun,$ $ 60.0\,\Msun]$, the Lense-Thirring precession has negligible contribution to the total advance of periastron.
The contribution to $\dot{\omega}_{\mathrm{SO}}$ in equation~\ref{omega_dot_SO} from a rotationally-induced quadrupole moment ${\dot{\omega}}_Q$ for a NS companion \cite{Wex_1995} or a BH counterpart \cite{Wex_Kopeikin_98} is
\begin{equation}
    \dot{\omega}_{Q} \sim 10^{-3} \, \dot{\omega}_{\mathrm{LT}}.  \label{omd_Q}
\end{equation}
This contribution is also negligibly small. 

In the alternative case of a massive WD, contributions to the $\dot\omega_{\mathrm{SO}}$ value from the Lense-Thirring precession and the quadrupole moment are of similar magnitude to each other \cite{Krishnan_2020}. The upper mass limit for a rigidly rotating WD is about $1.47 \, \Msun$ \cite{2005A&A...435..967Y}, so we consider an uniformly rotating $1.4\, \Msun$ WD for the companion of \E. The typical spin value for such a massive WD with extreme rotation is $S_\mathrm{c} \sim 0.6 \times 10^{50} \rm \,g\, cm^{2}\,s^{-1}$ \cite{Boshkayev_2017} and assuming the minimum pulsar mass of $1.1 \, \Msun$ (see above), we used equations \ref{omega_dot_LT}, \ref{B_c}, and \ref{f_c} above to calculate the maximum contribution from the Lense-Thirring precession:
\begin{equation}
    |{\dot\omega}_{\mathrm{LT}}|^\mathrm{max}/{{\dot\omega}_M}(m_\mathrm{p} = 1.1\, {\Msun},m_\mathrm{c} = 1.4\, {\Msun}) = 0.0016.
\end{equation}
Using $Q = 0.6\times{10^{49}}\:\mathrm{g}\:\mathrm{cm^{2}}$ for a rotating WD \cite{Boshkayev_2014}, we calculate
\begin{equation}
    {\dot\omega}_{Q}/{{\dot\omega}_M}(m_\mathrm{p} = 1.1\, {\Msun},m_\mathrm{c} = 1.4\, {\Msun}) = 0.0039. \label{Q_WD}
\end{equation}
This is too low for the observed advance of periastron to be explained by a (uniformly rotating) WD, even if it is rotating at the mass-shedding limit. We conclude that $\dot\omega_{\mathrm{obs}}$ is dominated by the leading order GR contribution (equation~\ref{omdot_rel}), and thus the inferred mass of the companion ($m_\mathrm{c} > 2.09\, \rm M_{\odot}$) is too large for it to be a massive WD.

\par
To summarize, there is negligible contribution to the rate of advance of periastron from the proper motion of the binary or the presence of a possible third body near the system. Over the whole range of companion masses investigated, the maximum possible change in the longitude of periastron due to Lense-Thirring precession, $\dot\omega_{\mathrm{LT}}$ (Figure \ref{fig:omega_dot_LT_variation}), is more than a factor of 100 smaller than the observed rate of advance of periastron (Table \ref{tab:ephemeris}). Combining this with equation~\ref{omdot_1}, we conclude that the observed change in the longitude of periastron of the pulsar is relativistic, and the total mass of the system can be approximated by the leading-order periastron advance term using equation~\ref{eq:Mtot}. 
\par
Figure \ref{fig:omega_dot_LT_variation} shows the constraint on the total mass of the system given by $\dot\omega_{\mathrm{obs}} = \dot\omega_{M}$. For any combination of pulsar and companion mass ($m_\mathrm{p}, m_\mathrm{c}$), the maximum possible contribution from Lense-Thirring precession is always less than 1\% of the observed rate of advance of periastron, and is therefore negligible.

\subsubsection*{Contributions to $\dot{P}_\mathrm{b,obs}$}

The observed variation in the orbital period of a binary pulsar ($\dot{P}_\mathrm{b,obs}$) can be expressed as \cite{Handbook_Lorimer_Kramer}:
\begin{equation}
    \dot{P}_\mathrm{b,obs} = \dot{P}_\mathrm{b,int} + \dot{P}_\mathrm{b,GC} + \dot{P}_\mathrm{b,Shk} + \dot{P}_\mathrm{b,Gal}  \label{Pbdot},
\end{equation}
where $\dot{P}_\mathrm{b,int}$ is the intrinsic orbital period derivative and $\dot{P}_\mathrm{b,GC}$ is the contribution from the line-of-sight acceleration of the system in the gravitational field of the cluster ($A$). The third and the fourth terms in this equation are the change in the orbital period of the binary due to the Shklovskii effect \cite{Shklovskii_1970} and the difference between the Galactic accelerations of the binary and the Solar System.
\\
The observed orbital period derivative (Table \ref{tab:ephemeris}) is $\dot{P}_\mathrm{b}\, = \, (18.1 \, \pm\, 5.6)\, \times 10^{-12} \,\mathrm{s}\,\mathrm{s}^{-1}$. The change in the orbital period from the Shklovskii effect, associated with the proper motion of the system, is given by:
\begin{equation}
    \dot{P}_\mathrm{b,Shk} = \frac{{\mu^2}d}{c}P_\mathrm{b} = 9.02 \times 10^{-14}\,\mathrm{s}\,\mathrm{s}^{-1} \label{Pbdot_Shkval}, 
\end{equation}
where $d$ is the distance between the Earth and the pulsar. Here we use the cluster distance estimate of \cite{2022ApJ...934..150L}, 11.66~kpc. With the available timing baseline, we cannot measure the proper motion of the binary so have used the proper motion of the cluster (see above). If the binary is gravitationally bound to the GC, the proper motion of the pulsar cannot differ much from the proper motion of the cluster; the same applies for $\dot{P}_\mathrm{b,Shk}$.
\\
We calculate the contributions to $\dot{P}_\mathrm{b,obs}$ from the Galactic acceleration using a Milky Way potential \cite{McMillan_2017}, finding: 
\begin{equation}
    \dot{P}_\mathrm{b,Gal} = -2.52 \times 10^{-14}\,\mathrm{s}\,\mathrm{s}^{-1} \label{Pbdot_Galval}.
\end{equation}
The sum of $\dot{P}_\mathrm{b,Shk}$ and  $\dot{P}_\mathrm{b,Gal}$ is almost 300 times smaller than $\dot{P}_\mathrm{b,obs}$, which implies that the latter is dominated by the effect of $A$:
\begin{equation}
    \dot{P}_\mathrm{b,obs} \simeq \dot{P}_\mathrm{b,GC} = \frac{A}{c} P_{\mathrm{b}}  \label{Pbdot_GC}.
\end{equation}
We conclude that $\dot{P}_\mathrm{b,obs}$ depends only on $A$.

\subsubsection*{Contributions to $\dot{x}_{\mathrm{obs}}$ from Lense-Thirring precession}
The observed change in the projected semi-major axis of the binary pulsar ($\dot{x}_{\mathrm{obs}}$) could be caused by a physical change in the orbit of the pulsar, a temporal change in the angle of inclination of the system, or a combination of both. While the semi-major axis of the pulsar is unaffected by the Lense-Thirring effect, precession of the binary orbit could potentially be observable as a slow change in the angle of inclination of the system, caused by the spins of the binary components. As discussed above, the spin angular momentum of the pulsar is about an order of magnitude smaller than the maximum possible value of the spin angular momentum of the companion so we neglect it in further analysis; however, the spin of the companion could lead to a secular change in the projected semi-major axis depending on its alignment. 

The change in the inclination angle as an effect of the Lense-Thirring precession of the compact companion is, to the leading order \cite{Damour_Schafer_1988,Damour_Taylor_92}:
\begin{equation}
  \left(\frac{di}{dt}\right)_{\mathrm{LT}}
  = {\frac{4 + 3{m_\mathrm{p}}/{m_\mathrm{c}}}{2{c^2}M(1-e^2)^{3/2}}}\:
    {{n}^2_\mathrm{b}}({\mathbf{S}}_\mathrm{c}\cdot{\hat{\mathbf{i}}}) 
  = {\frac{Gm_\mathrm{c}(3M + m_\mathrm{c})}{2{c^3}M(1-e^2)^{3/2}}}\:
    {{n}^2_\mathrm{b}}\,{\chi_\mathrm{c}}{\sin{\vartheta}}\,{\sin{\mathrm{\phi}_{\mathrm{SO}}}} \,.
\label{didt_LT}
\end{equation}
Secular evolution in the angle of the inclination would occur if the spin of the companion has a non-negligible component along the direction of the ascending node. This, in turn, would change the observed projected semi-major axis:
\begin{equation}
    {\dot{x}}_{\mathrm{LT}} = {x}\,{\cot{i}}\left(\frac{di}{dt}\right)_\mathrm{LT} \,.
\label{xdot_LT}
\end{equation}
To estimate the maximal contribution to ${\dot{x}}_{\mathrm{LT}}$, we consider a pulsar of $1.2\,\Msun$ and a light BH companion with a mass of $2.7\,\Msun$. For this mass pair, the maximum observable value of the variation in the projected semi-major axis of the pulsar is $\dot{x} = (1.7 \times 10^{-13})\,\chi_\mathrm{c}$. For a Kerr BH $\chi_\mathrm{c} \le 1$. If this BH is the result of a NS binary merger (irrespective of the collapse time), the maximum value of the dimensionless spin of the companion is theoretically expected to lie in the range $\chi_\mathrm{c} \in [0.6,0.875]$ \cite{Bernuzzi_2020}.

\par
We left $\dot{x}$ as a free parameter in our radio timing analysis to determine the maximum expected value. We calculated this using the DDGR model, due to the large expected correlation between $\gamma_{\rm E}$ and $\dot{x}$ \cite{2019MNRAS.490.3860R}; this method has been used in previous measurements of $\dot{x}$ \cite{Krishnan_2020}. Assuming a range of companion masses ($2.3\,\Msun \leq m_\mathrm{c} \leq 2.7\,\Msun$) we obtain $\dot{x} \sim (4\,\pm\,2) \times 10^{-13}$. This is not significant, and the 1-$\sigma$ uncertainty is larger than our estimated upper limit for $\dot{x}_{\mathrm{LT}}$.
\par

\par
In addition to the precession of the orbit due to spin-orbit coupling, there are several other geometric and physical effects that could influence the observed change in the projected semi-major axis: these include the proper motion of the binary, the emission of gravitational waves or mass loss from the system, a change in the aberration or Doppler shift, and the presence of a hypothetical third body around the binary. However, the contributions from any of these terms are at least an order of magnitude smaller than the contribution from the Lense-Thirring effect, and are hence neglected in our analysis.

\newpage

\newpage

\newpage

\newpage

\newpage

\newpage

\newpage

\newpage

\newpage

\newpage

\newpage

\begin{table}[H]
\caption{\textbf{Radio timing observations of {\E}.} Shapiro delay campaign observations (indicated by an asterisk) covered orbital phase ranges of 0.21--0.23, 0.61--0.63, 0.88--0.90 and 0.98--0.034.}
\label{tab:observations}
\begin{tabular}{rrllrrr}
\hline
\hline
\multicolumn{1}{l}{Start time} & \multicolumn{1}{l}{Start time} & Telescope & Backend & \multicolumn{1}{l}{Duration} & \multicolumn{1}{l}{Centre frequency,} & \multicolumn{1}{l}{Bandwidth,}        \\
\multicolumn{1}{l}{(UTC)}      & \multicolumn{1}{l}{(MJD)}      &           &         & \multicolumn{1}{l}{(s)}      & \multicolumn{1}{l}{$f_\mathrm{ctr}$ (MHz)}       & \multicolumn{1}{l}{$\Delta{f}$ (MHz)} \\
\hline
27 Dec. 2005                   & 53731.127                      & GBT       & SPIGOT  & 14160                        & 820.024                               & 50                                    \\
17 Feb. 2006                   & 53783.992                      & GBT       & SPIGOT  & 9135                         & 820.024                               & 50                                    \\
18 Apr. 2006                   & 53843.809                      & GBT       & SPIGOT  & 5098                         & 820.024                               & 50                                    \\
19 Apr. 2006                   & 53844.820                      & GBT       & SPIGOT  & 11037                        & 820.024                               & 50                                    \\
2 Jun. 2006                    & 53888.728                      & GBT       & SPIGOT  & 7197                         & 820.024                               & 50                                    \\
11 Aug. 2006                   & 53958.489                      & GBT       & SPIGOT  & 7559                         & 1949.609                              & 600                                   \\
15 Jan. 2021                   & 59229.626                      & MeerKAT   & PTUSE   & 14400                        & 1283.896                              & 856                                   \\
8 May. 2021                    & 59342.278                      & MeerKAT   & PTUSE   & 7200                         & 1283.896                              & 856                                   \\
8 May. 2021                    & 59342.586                      & MeerKAT   & PTUSE   & 7200                         & 1283.896                              & 856                                   \\
21 May. 2021                   & 59355.253                      & MeerKAT   & PTUSE   & 7200                         & 815.934                               & 544                                   \\
21 May. 2021                   & 59355.628                      & MeerKAT   & PTUSE   & 7200                         & 815.934                               & 544                                   \\
24 May. 2021                   & 59358.242                      & MeerKAT   & PTUSE   & 7200                         & 815.934                               & 544                                   \\
26 May. 2021                   & 59360.607                      & MeerKAT   & PTUSE   & 7200                         & 815.734                               & 544                                   \\
3 Jul. 2021                    & 59398.485                      & MeerKAT   & PTUSE   & 7200                         & 815.734                               & 544                                   \\
17 Aug. 2021                   & 59443.233                      & MeerKAT   & PTUSE   & 7200                         & 815.734                               & 544                                   \\
1 Sep. 2021                    & 59458.067                      & MeerKAT   & PTUSE   & 7200                         & 815.734                               & 544                                   \\
30 Sep. 2021                   & 59487.942                      & MeerKAT   & PTUSE   & 7200                         & 815.734                               & 544                                   \\
7 Nov. 2021                    & 59525.075                      & MeerKAT   & PTUSE   & 10800                        & 815.734                               & 544                                   \\
9 Nov. 2021                    & 59527.919                      & MeerKAT   & PTUSE   & 7200                         & 815.734                               & 544                                   \\
6 Feb. 2022                    & 59616.544                      & MeerKAT   & PTUSE   & 7200                         & 815.734                               & 544                                   \\
12 Apr. 2022                   & 59681.483                      & MeerKAT   & PTUSE   & 7200                         & 815.734                               & 544                                   \\
27 May. 2022                   & 59726.630                      & MeerKAT   & PTUSE   & 7200                         & 815.734                               & 544                                   \\
29 May. 2022                   & 59728.322                      & MeerKAT   & PTUSE   & 7200                         & 815.734                               & 544                                   \\
2 Jun. 2022                    & 59732.609                      & MeerKAT   & PTUSE   & 7200                         & 815.734                               & 544                                   \\
27 Jul. 2022$^{*}$                   & 59787.380                      & MeerKAT   & PTUSE   & 10800                        & 815.934                               & 544                                   \\
30 Jul. 2022$^{*}$                    & 59790.380                      & MeerKAT   & PTUSE   & 10800                        & 815.934                               & 544                                   \\
1 Aug. 2022$^{*}$                     & 59792.401                      & MeerKAT   & PTUSE   & 10800                        & 815.934                               & 544                                   \\
2 Aug. 2022$^{*}$                     & 59793.161                      & MeerKAT   & PTUSE   & 9900                         & 815.934                               & 544                                   \\
2 Aug. 2022$^{*}$                     & 59793.285                      & MeerKAT   & PTUSE   & 9900                         & 815.934                               & 544                                   \\
2 Aug. 2022$^{*}$                     & 59793.409                      & MeerKAT   & PTUSE   & 9900                         & 815.934                               & 544                                   \\               
\hline
\end{tabular}
\end{table}

\newpage   

\begin{table}[H]
       \caption{\textbf{Masses and references of BH/NS components within or near the lower mass gap.} For GW sources, the error bars are 90\% symmetric credible intervals. For GW170817, we followed \cite{Shibata_2019,2017PhRvD..95b4029D} and estimated the upper mass limit assuming a remnant mass of 90\% of sum of two merging NS components minus $0.04\;M_\odot$ of baryonic ejecta mass. For the radio pulsars, see the individual references in the table for a description of how error bars on masses are calculated in more detail.} 
       \begin{center}
   \begin{tabular}{lll}
       \toprule[0.1ex]
       \midrule[0.1ex]
       \multicolumn{2}{l}{Gravitational-wave mergers:}\\
       \midrule[0.1ex]
       GW190917 & $2.1^{+1.1}_{-0.4}\;M_\odot$ & \cite{2021arXiv211103606T}\\
       GW190814 & $2.59^{+0.08}_{-0.09}\;M_\odot$ & \cite{2020ApJ...896L..44A}\\
       GW200210 & $2.83^{+0.47}_{-0.42}\;M_\odot$ & \cite{2021arXiv211103606T}\\
       \midrule[0.1ex] 
       \multicolumn{2}{l}{Remnant mass of double NS merger:}\\
       \midrule[0.1ex]
       GW170817 & $2.46^{+0.23}_{-0.14}\;M_\odot$ & \cite{Shibata_2019,2017PhRvD..95b4029D}\\
       \midrule[0.1ex] 
       \multicolumn{2}{l}{Radio pulsars:} \\
       \midrule[0.1ex]
       PSR~J0514$-$4002E$^{\ast}$ & $2.31^{+0.41}_{-0.22}\;M_\odot$ & This work \\
       PSR~J0348+0432 & $2.01\pm 0.04\;M_\odot$ & \cite{2013Sci...340..448A} \\
       PSR~J0740+6620 & $2.08\pm 0.07\;M_\odot$ & \cite{2021ApJ...915L..12F} \\
       \bottomrule[0.1ex] 
       \multicolumn{2}{l}{$^*$Mass of the companion star.}
    \end{tabular}
    \label{tab:mass-gap-references}
   \end{center} 
\end{table}

\end{document}